\documentstyle[prl,aps,epsf]{revtex}
\begin{document}
 
\twocolumn[\hsize\textwidth\columnwidth\hsize\csname
@twocolumnfalse\endcsname
\title{A first-principles approach to electrical transport in atomic-scale 
nanostructures}
\author{J. J. Palacios$^*$, A. J. P\'erez-Jim\'enez$^{**}$, 
E. Louis$^*$, E. SanFabi\'an$^{**}$, and J. A. Verg\'es$^{***}$}
\address{$^*$Departamento de F\'{\i}sica Aplicada,
Universidad de Alicante, San Vicente del Raspeig, Alicante 03690, Spain}
\address{$^{**}$Departamento de Qu\'{\i}mica-F\'{\i}sica,
Universidad de Alicante, San Vicente del Raspeig, Alicante 03690, Spain}
\address{$^{***}$Instituto de Ciencia de Materiales de Madrid (CSIC),
Cantoblanco, Madrid 28049, Spain.}
\date{\today}
\maketitle
  
\begin{abstract} 
We present a first-principles numerical implementation of Landauer formalism
for electrical transport in nanostructures characterized down to the atomic
level.  The novelty and interest of our method lies essentially on two facts.
First of all, it makes use of the versatile Gaussian98 code, which is widely
used within the quantum chemistry community. Secondly, it incorporates the
semi-infinite electrodes in a very generic and efficient way by means of Bethe
lattices.  We name this method the Gaussian Embedded Cluster Method (GECM). In
order to make contact with other proposed implementations, we illustrate our
technique by calculating the conductance in some well-studied systems such as
metallic (Al and Au) nanocontacts and C-atom chains connected to metallic (Al
and Au) electrodes.  In the case of Al nanocontacts the conductance turns out
to be quite dependent on the detailed atomic arrangement. On the contrary,
the conductance in Au nanocontacts presents quite universal features.
In the case of C chains, where the self-consistency
guarantees the local charge transfer and 
the correct alignment of the molecular and
electrode levels,  we find that the conductance oscillates with
the number of atoms in the chain
regardless of the type of electrode. However, for short
chains and Al electrodes the even-odd periodicity is reversed at equilibrium
bond distances.
\end{abstract} 

\vskip2pc]
 
\section{Introduction}

Molecular- and atomic-scale electronic devices are attracting an ever
increasing interest due to the impact they are expected to make in the world of
Nanotechnology. The number of experimental and theoretical works in this
particular area of research, generically known as molecular
electronics\cite{JGA01}, is growing exponentially.  The design of devices at
the molecular and even atomic scale poses new challenges which require new
theoretical and experimental techniques to be developed.  Scanning Tunneling
Microscopy (STM) is probably the pioneer of the experimental techniques in this
research area. It can be used not only in the tunneling regime to image
adsorbates\cite{Jo95,Pa00}, but also in the contact regime to build few-atom
nanoscopic contacts\cite{Ag93}.  STM can also be used to investigate the
electrical properties of nanotubes\cite{De99} and DNA molecules\cite{Go00} with
one or both of their ends attached  to a suitable electrode.  
In addition to STM,
mechanically controllable break junctions have also revealed themselves as
powerful tools to study electrical transport in metallic 
nanobridges\cite{Sc98} or individual molecules\cite{Re97,Ke99}.

The basics to calculate the zero-bias
conductance $G$ of a nanoscale contact had
been established by Landauer's in his pioneering work\cite{Da95} long before
these systems were commonplace. In Landauer's formalism $G$ 
is simply given by the quantum mechanical transmission of
the electrons around the Fermi energy\cite{MW92}. 
The value of this transmission is
essentially determined by the region where the number of channels available for
conduction is the smallest. In molecular or atomic-scale
nanocontacts the region of relevance is the molecule and/or the few atoms 
forming
the nanoscopic bridge between electrodes.  The transmission is thus strongly
dependent on the particular molecule, the detailed atomic arrangement of the
electrodes in the contact region, and the chemical nature of them. 
Knowing the atomic arrangement of the electrodes 
or the way the molecule binds to the electrodes is, however, 
a major problem in itself.  Furthermore, even if these important details 
were known, implementing Landauer's formalism still requires to
know the electronic structure and this is a formidable task as well.  

Calculations based on tight-binding or semi-empirical 
models\cite{Jo98,Cu98,BKT99}
have been, and still are, very popular since these models
capture the atomic-scale character in some detail and are easy to implement.
However, they do not allow for structural relaxations to be performed. Most
importantly, these simple models, in general, do not yield correct values for
the local electronic charges. In other words, the chemical potential is 
not uniform across the entire system in equilibrium.  
While imposing local charge
neutrality is a straightfoward improvement on these models for metallic
nanoconstrictions\cite{Cu98,BKT99}, there does not exist any simple
modification in the case of more complex systems like metal-molecule-metal
heterostructures\cite{PS01}. 
A way around this problem is to perform self-consistent
first-principles calculations which, at least at a mean-field level, guarantee
the uniformity of the chemical potential. However, most numerical
implementations commonly used to carry out {\it ab-initio} electronic
calculations are either restricted to finite systems, such as the Gaussian
code\cite{Gaussian}, or require the infinite system to be periodic such as the
SIESTA code\cite{SOAS99}.  None of these methods are suitable to address the
systems here studied which are both infinite and non-periodic.  Finally, a
perhaps more serious difficulty is the intrinsic non-equilibrium character of
electrical transport. 

In recent years several proposals have appeared to tackle this
problem\cite{La95,HT95,TGW01a,KAT01,BT01,XDR01}.  Most of all are based
upon density functional (DF) theory. In addition to the well-known virtues of
the DF theory, it presents the additional advantange that Landauer's
theoretical framework does not need to be modified since DF is still a
single-particle description of the many-body problem.  In the pioneering works
of Lang and co-workers, Tsukada and co-workers, and Guo and
co-workers\cite{La95,HT95,LA98,Ta98,Wa97} the electrodes were described within
the jellium approximation. Jellium models are still used
today\cite{KAT01,LA00,DK01}; for they are convenient in one way: They provide
featureless contacts which represent generic situations.  As mentioned below
this is one major feature of our model too.  However the jellium model presents
serious drawbacks.  How can the differences observed experimentally between,
for instance, Al and Au electrodes, be taken into account by means of a
jellium?. Furthermore, this approach is not satisfactory when one is trying to
describe, for instance, STM experiments where the detailed atomic structure of
the tip determines, to a large extent, whether or not the STM can resolve the
topography or molecular structure of the adsorbate.  Recent approaches, which
essentially differ only on the numerical implementation, intend to 
incorporate the atomic structure of the electrodes in the DF
calculation\cite{TGW01a,BT01,XDR01,TGW01b,MW01}.  It is
pertinent noting here that in most of these studies a periodic structure beyond
a given point within the leads is assumed.  Efficient transfer matrix
techniques\cite{GT83,LL84} make this  reasonable assumption tractable, but it
forces one to consider a very specific type of lead (typically a
finite-section wire\cite{TGW01a,BT01,TGW01b,MW01} or 
an infinite surface\cite{XDR01}). Using a jellium model for the 
electrodes is harmless but it lacks the minimum atomic detail which 
is crucial to describe, e.g., the contact with molecules. However,  
employing  well-defined specific electrodes
is not desirable either since their own electronic structure can interfere
with the interpretation of the results. For instance, the
appearance of gaps in the conductance  close
to the Fermi energy for perfectly conducting systems such as Au chains
can only be considered
an artifact due to the unphysical electrode model\cite{BT01,MW01}. 
Furthermore, actual nanocontacts are not expected to have high symmetry.  
As explained below these difficulties are circumvented in our method.

Recently\cite{PP01}, we have presented an alternative to the {\em ab-initio}
methods above mentioned.  Close in spirit to those presented in Refs. 
\onlinecite{TGW01a,BT01}, the main differences and advantages with respect to
them are the use of the standard
Gaussian98 code to carry out the DF calculation of the relevant transport
region and the description of the electrodes bulk by means of appropriate Bethe
lattices\cite{MV90,JY74}.  The Gaussian98 code provides a versatile method to
perfom first-principles calculations of clusters, incorporating the major
advancements in the field in terms of functionals, basis sets,
pseudopotentials, etc.. On the other hand, the Bethe lattices are two-fold
convenient: (i) They reproduce the essential features of the bulk density of
states and (ii) their directional self-energies can be easily calculated (see
Appendix).  In Ref.\ \onlinecite{PP01} these ideas were applied to investigate
electrical transport of a C$_{60}$ molecule in between Al electrodes.
Here we describe in detail an
improvement to our previous approach\cite{PP01} which can be summarized in that
we now incorporate self-consistently 
the semi-infinite electrodes into the conductance calculation within the
Gaussian98 code.  This
requires working with Green's functions from the very start. The method has some
resemblance with the cluster Bethe lattice method developed to investigate the
physical properties (electronic structure, phonons, etc.) of disordered
systems\cite{JY74}. We illustrate the possibilities of our method by
investigating electronic transport in Al and Au nanocontacts. These systems
have been the subject of extensive studies in the past by means of
tight-binding methods\cite{Cu98,BKT99} and, more recently, by means of {\em
ab-initio} methods\cite{KAT01,BT01,MW01}. We show that, even in these systems
where charge transfer is apparently unessential, tight-binding methods may fail
to provide  a correct quantitative picture. It turns out that the conductance
in the case of Al nanocontacts is strongly dependent on the
detailed atomic structure. In the case of Au nanocontacts, on the contrary,
the results are more universal as confirmed by experiments.
Next, we choose a system with somewhat appealing features: Carbon-atom chains.
As shown in Ref.\ \onlinecite{LA98} these chains, when contacted by Al
electrodes, exhibit a conductance that oscillates with the number of atoms in
the chain.  Here we address this problem by taking proper account of the 
binding to the electrodes and investigate how the results depend on the type
of electrode (Al or Au).

The rest of the paper is organized as follows. In Section II we discuss
the main characteristics of our method. 
Section III is devoted to an extensive discussion of the results.
Finally, we end the paper by summarising the main
features of the method and the most remarkable results (Section IV).

\section{The Gaussian Embedded Cluster Method (GECM)}

\begin{figure}[tb]
\vspace{-0cm}
\centerline{\epsfxsize=6cm \epsfysize=6cm \epsfbox{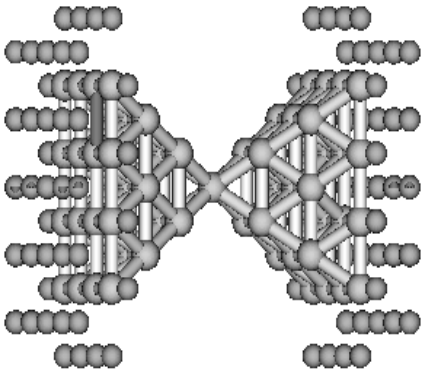}}
\caption{Schematic view of a cluster where phantom atoms from the Bethe 
lattices are shown.}
\label{bethe}
\end{figure}

In previous work\cite{PP01} we have presented a method  to study
transport in atomic-scale and molecular devices which is
based on standard quantum chemistry calculations with the Gaussian98
code\cite{Gaussian}. This scheme, which has been recently 
adopted by other groups
(see, e.g., Refs.\ \onlinecite{Cu01}), is taken here a step further.
A DF calculation of the region that includes 
the molecule or set of atoms forming the contact between 
electrodes and a significant part of the electrodes is performed
(see Fig. \ref{bethe}).  As far as transport is concerned,
the hamiltonian (or Fock matrix $\hat{F}$)
of this central cluster or supermolecule
contains the relevant information since it embraces the region with
the smallest number of channels for conduction. However,
 according to the usual theoretical transport
schemes\cite{Da95}, its associated Green's
functions are unsuitable for the evaluation of the current
(note that they simply have poles).
The retarded(advanced) Green's functions associated with  $\hat{F}$
needs to be extended to include the rest of the semi-infinite electrodes:
\begin{equation}
\hat G^{r(a)}(\epsilon)= (\epsilon\hat I\hat S  - 
 \hat F \pm i\delta)^{-1} \rightarrow 
[\epsilon\hat I \hat S -  \hat F - \hat\Sigma^{r(a)}(\epsilon) ]^{-1}.
\label{green}
\end{equation}

\noindent In this expression

\begin{equation}
\hat\Sigma^{r(a)}(\epsilon)=\hat\Sigma_R^{r(a)}(\epsilon) +
 \hat\Sigma_L^{r(a)}(\epsilon), 
\end{equation}

\noindent where $\hat\Sigma_R$($\hat\Sigma_L$) denotes a self-energy matrix
that accounts for the part of the right(left) semi-infinite electrode which has
not been included in the DF calculation. $\hat S$ is the overlap matrix and
$\hat I$ is the unity matrix.  The added self-energy matrices can only be
explicitly calculated in ideal situations, which, in principle, limits the
desired applicability of the above procedure. For instance, in Refs.\
\onlinecite{BT01,MW01} the authors  consider finite-section wires as
electrodes.  As a result of this choice gaps appear
in the conductance of otherwise perfectly
conducting central clusters.  

In order to overcome this type of problem, we
choose to describe the bulk electrode with a Bethe lattice tight-binding model
with the appropriate coordination numbers and parameters (see Appendix).  The
advantage of choosing a Bethe lattice resides in that it reproduces fairly well
the bulk density of states of any metallic electrode, avoiding this way the
appearance of spurious results. In addition to this the self-energy matrices
that appear in Eq. \ref{green} can be calculated iteratively in a simple way
(see Appendix for more details).  For each atom of the outer planes of the
cluster, we choose to add a branch of the Cayley tree in the direction of any
missing bulk atom (including those missing in the same plane).  In Fig.\
\ref{bethe} the directions in which  branches are added are indicated by 
smaller atoms which represent the first atom of the branch in that
direction. Assuming that the most important structural details of the electrode
are included in the central
cluster, the Bethe lattices should have no other relevance
than that of introducing a featureless reservoir.

In our present approach the self-consistent process does not stop once the
finite central cluster has been solved. Instead, we reformulate 
the Gaussian98 code
to proceed with the self-consistency of the now infinite
system. More specifically, once self-consistency for the finite cluster
has been almost attained, we calculate the Green's function as explained
above. Then, the density matrix is obtained from
the Green's function according to
\begin{equation}
\hat n=-\frac{1}{\pi}\int_{-\infty}^{\epsilon_F}{\rm Im}
\left [ \hat G^r(\epsilon)\right ]{\rm d}\epsilon,
\label{eqn:nab}
\end{equation}
\noindent where $\epsilon_F$ is the Fermi level fixed by the condition of
overall charge neutrality in the cluster.  The integral in Eq. (\ref{eqn:nab})
is calculated along a contour in the complex plane as explained in Refs.\
\cite{TGW01a,BT01,XDR01} with an efficient automatic numerical integration
scheme of P\'erez-Jord\'a {\em et al.}\cite{PJ92}.  The density matrix is now
used to recalculate the matrix elements of the Fock's operator and the process
is repeated until self-consistency is achieved.  We note that in this method
the standard eigenvalue problem, inherent to the Gaussian98 code, is replaced
by the calculation of Green's functions.  In the end the Green's functions
describe an infinite system in a more consistent way than in the method
discussed previously by us\cite{PP01} since it effectively removes finite-size
effects in the self-consistency. It is interesting to note that the
applicability of this approach, which we hereafter name the Gaussian Embedded
Cluster Method (GECM), goes well beyond the present study. In fact it could be
a powerful tool whenever an infinite media has to be described (for instance
adsorption of molecules on solid or liquid surfaces).

The conductance can now be simply
calculated through the expression\cite{Da95}
\begin{equation}
G=\frac{2e^2}{h}{\rm Tr}[\hat T],
\label{g}
\end{equation}
where  Tr denotes the trace over all the orbitals of the cluster and 
$\hat T$ is the transmission matrix which, in turn, is given by
\begin{equation}
\hat T = \hat\Gamma_L\hat G^r \hat\Gamma_R\hat G^a,
\end{equation}
where the matrices $\hat\Gamma_R$ and $\hat\Gamma_L$
are given by $i(\hat\Sigma^r_R-\hat\Sigma^a_R)$ and
$i(\hat\Sigma^r_L-\hat\Sigma^a_L)$, respectively. 
In order to single out the contribution of individual channels
to the current one can diagonalize the transmission matrix.
It turns out (see below) that only few channels give a
non-negligible contribution to the current. The symmetry of
each channel was identified by looking at its weight on
the atomic orbitals of the central atom of the constriction or the C chain.

Finally, it is worth pointing out that if  the current
in the finite-bias regime has to be obtained, one should simply integrate 
in energy
the above expression with appropriate Fermi distributions functions. 
Note, however, that a new definition of the density matrix generalized to 
non-equilibrium\cite{TGW01a,BT01,XDR01} needs to be used in the calculation
of the Green's functions. Apart from this, the Landauer-type 
expression (\ref{g}) remains valid as long as one does not give up
the single-particle description.
In this work we are concerned with basic and still open
aspects of transport in the systems studied and we will focus on the linear 
regime.

\section{Results}

For all the DF calculations we have used the Becke's three-parameter 
hybrid functional using the Lee, Yang and Parr correlation functional 
(B3LYP)\cite{Be93} together with the semilocal shape consistent 
pseudopotential (SCPP) and basis sets of Christiansen {\em et al.} 
\cite{CE85,CE86,CE90}. 
We have selected this combination of exchange-correlation functional and 
pseudopotential for two reasons: First, B3LYP is one of the most accurate
and certainly the most popular among the gradient-corrected 
exchange-correlation functionals; second, 
Christiansen {\em et al.} SCPP provide accurate results for a wide variety
of atoms while retaining the simplicity of a minimal basis set 
\cite{CE85,CE86,CE90,ST00}. 
Needless to say that there is no need to restrict the calculations to a
minimal basis set since the Bethe lattice can be constructed for any
basis set. It is only a matter of computational convenience that we
have done it so.  Nevertheless, in some  cases, we have checked that
better basis sets and the use of other exchange-correlation functionals do not 
modify the main conclusions of our work.

In all cases we have investigated the influence of
the number of electrode atoms  included in the DF calculation on the 
conductance. 
In general, the results do not vary qualitatively with the cluster size,
allowing us to extract some general conclusions. However, at
a quantitative level, this is not always
the case, particularly for Al nanocontacts. 
Finally we note that, although  most of the 
calculations were carried out taking the 
bulk inter-atomic distances for the electrode atoms, 
in some cases we investigated 
the effects of {\em ab-initio} relaxations.

\subsection{Metallic nanocontacts}
\begin{figure*}
\centerline {\epsfxsize=3cm \epsfysize=3cm
 \epsfbox{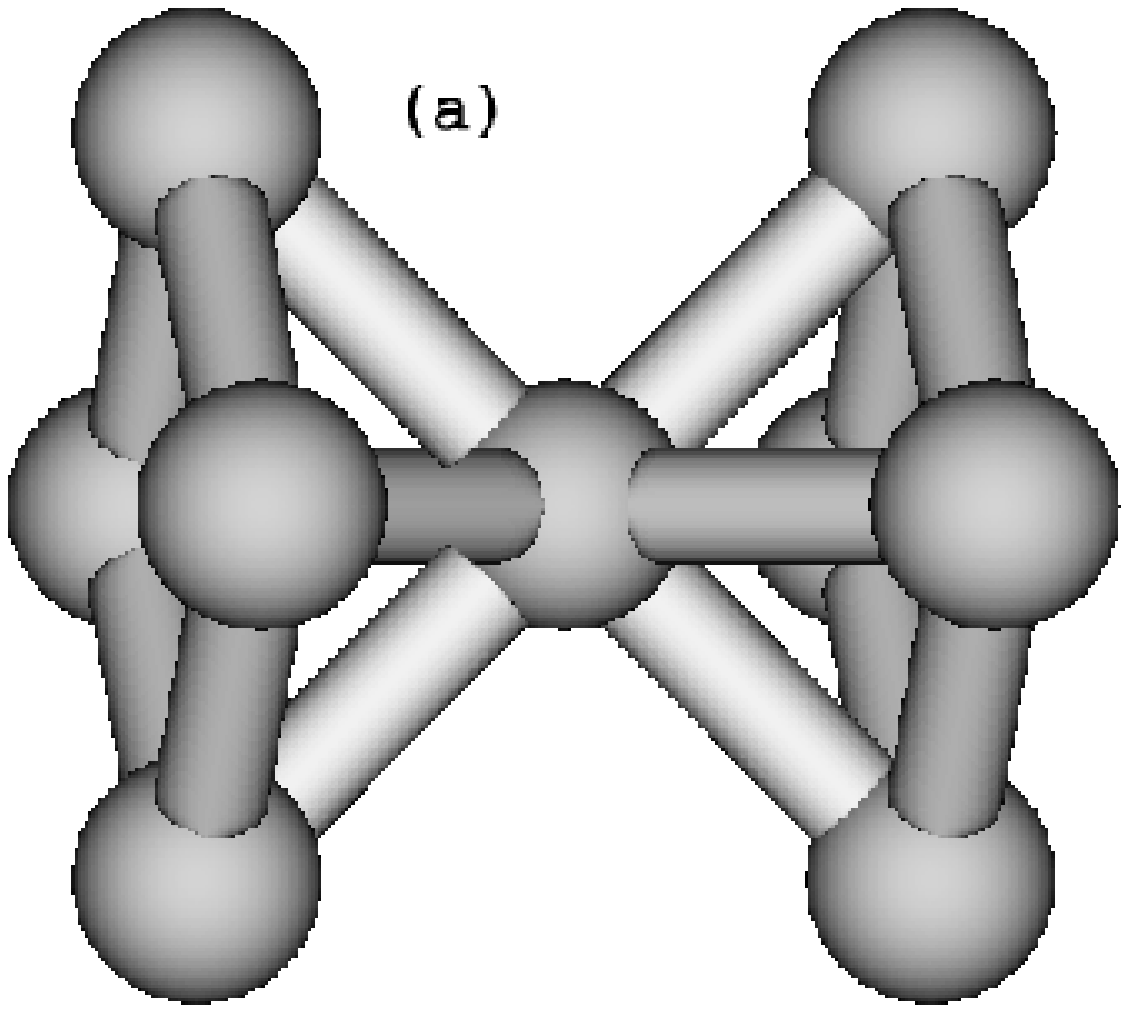}}
\centerline {\epsfxsize=3cm \epsfysize=3cm
 \epsfbox{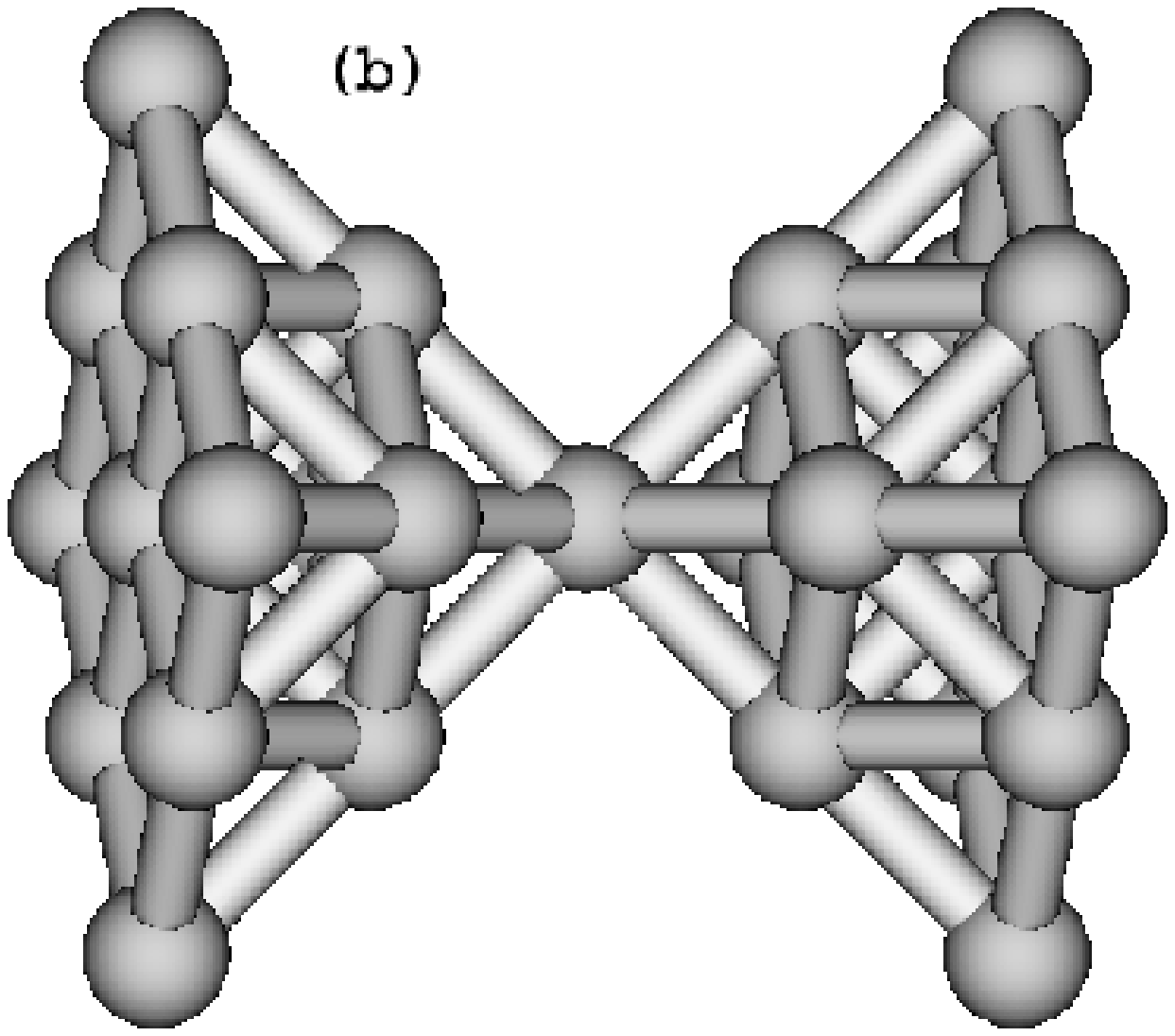}}
\centerline {\epsfxsize=3cm \epsfysize=3cm
 \epsfbox{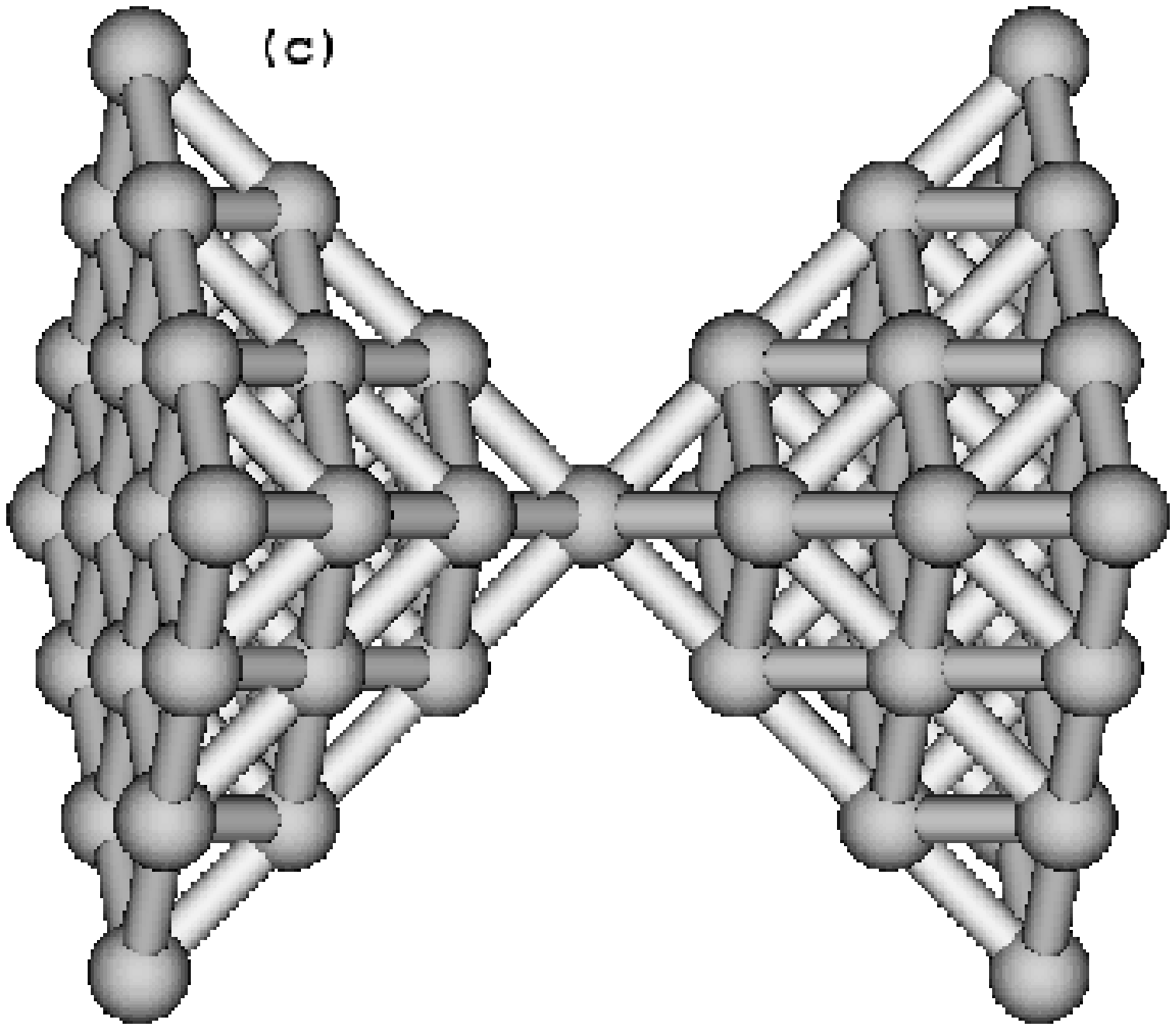}}
\vspace{0.0cm}
\caption{Atomic structure of the single-atom contact model 
considered in this work. 
The number of (001) planes increases by one from (a) to (c) in both electrodes 
increasing the size of the pyramids and the corresponding atomic detail
in the electrode bulk. Interatomic 
bulk distances have been considered for the whole cluster.}
\label{tip-atom-tip}
\end{figure*}

\begin{figure*}
\vspace{-0.8cm}
\centerline {\epsfxsize=3cm \epsfysize=3cm
 \epsfbox{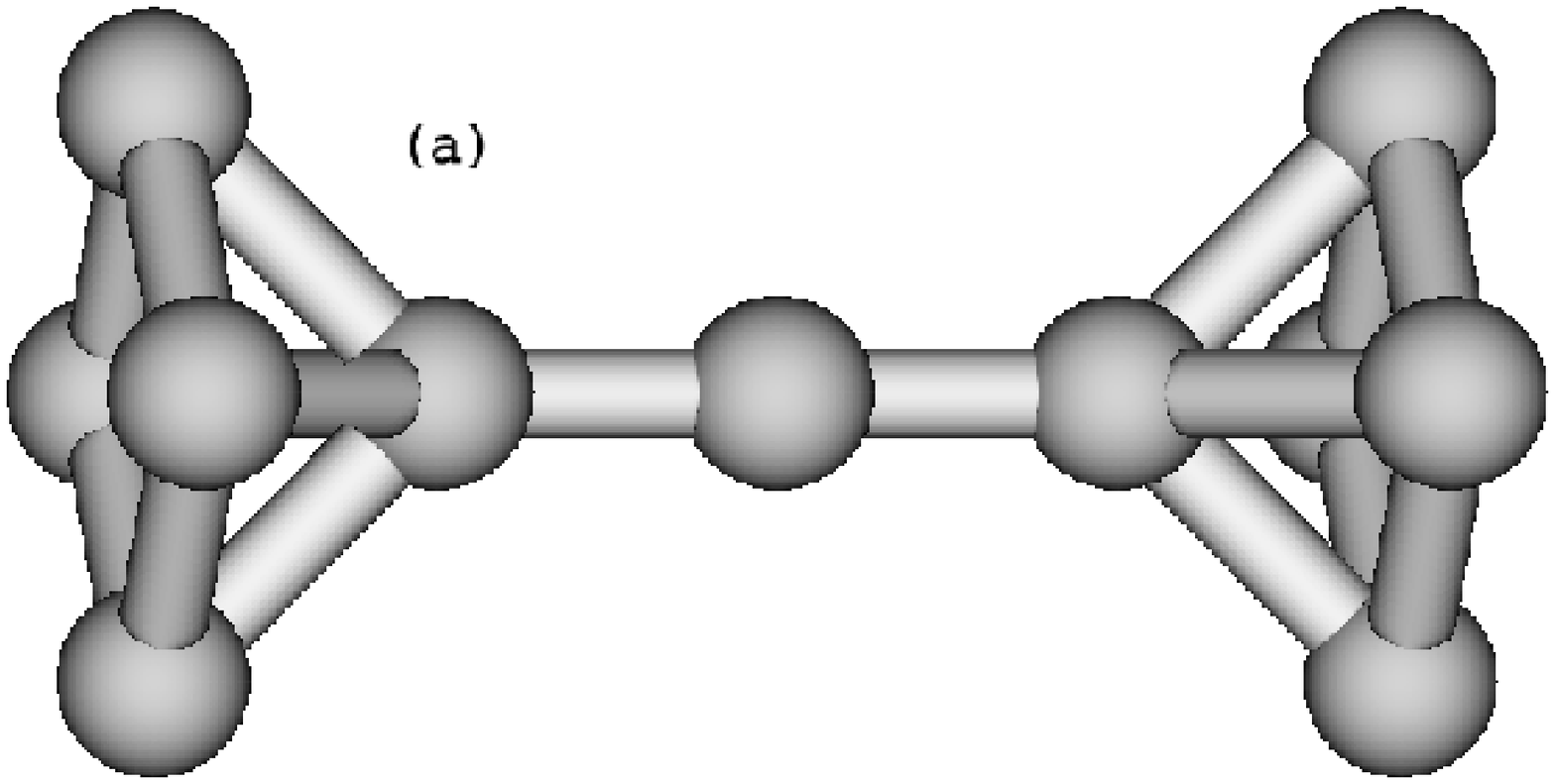}}
\vspace{-0.0cm}
\centerline {\epsfxsize=3cm \epsfysize=3cm
 \epsfbox{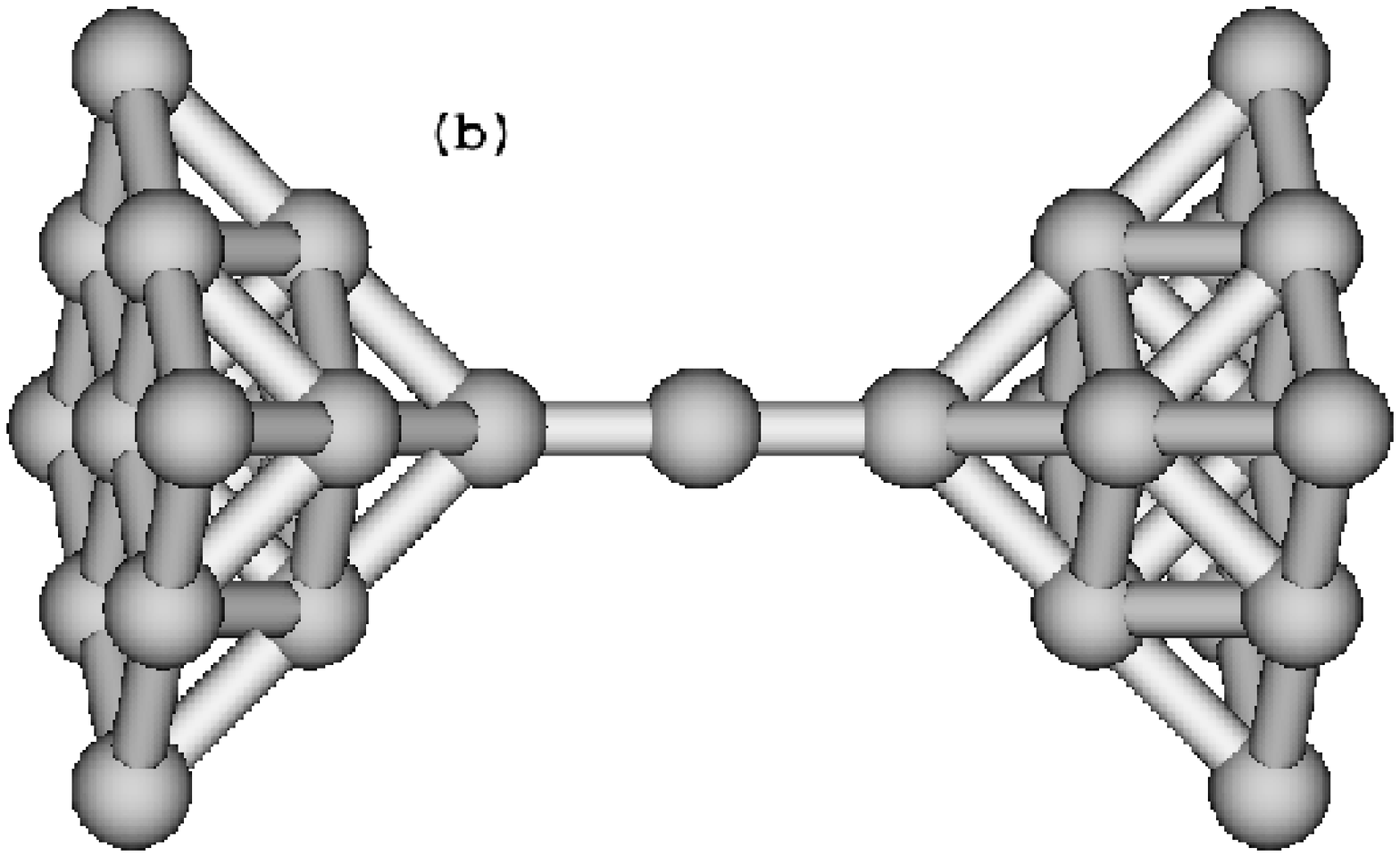}}
\vspace{-0.0cm}
\centerline {\epsfxsize=3cm \epsfysize=3cm
 \epsfbox{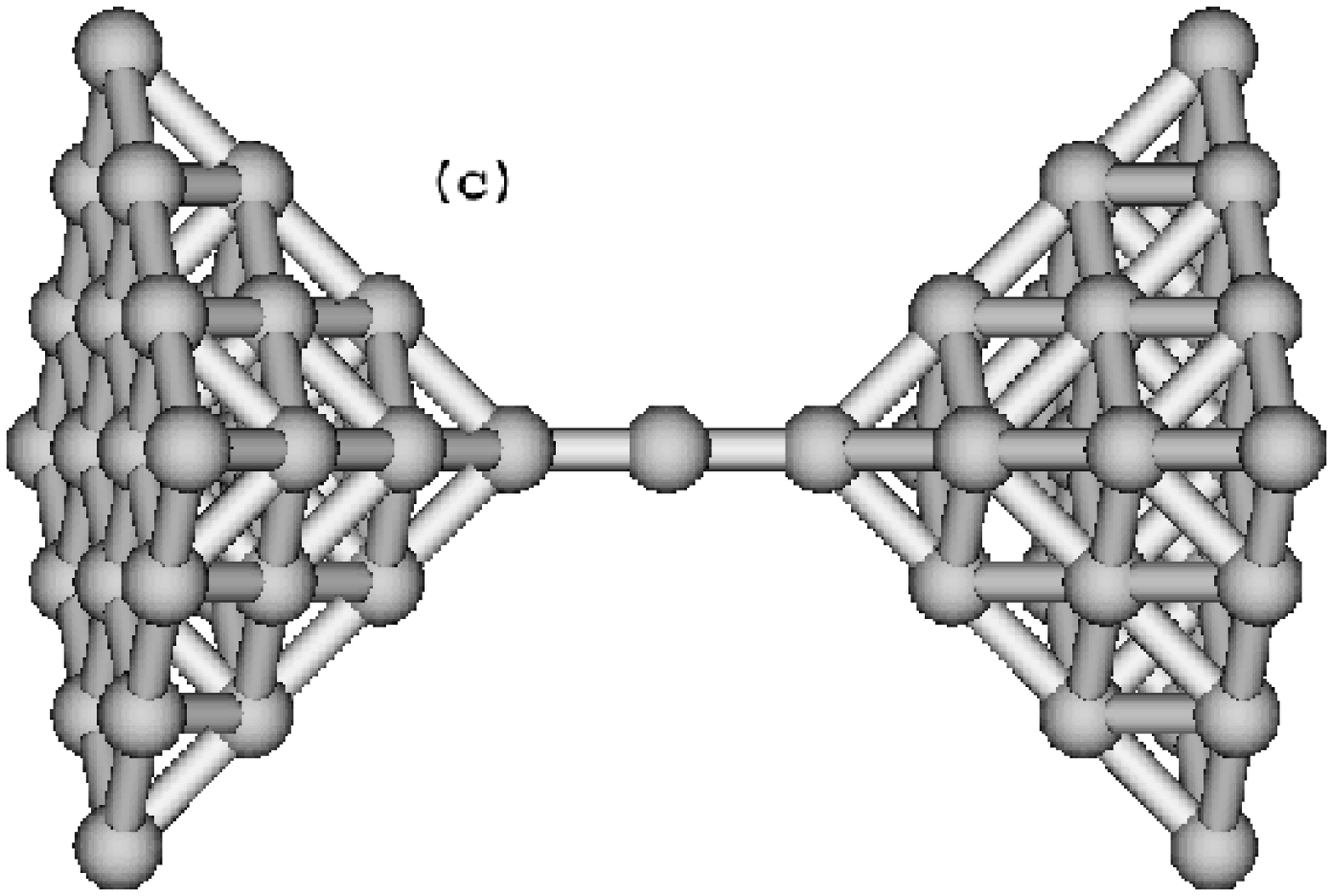}}
\caption{Atomic structure of the first 
atom-chain constriction model considered. 
The number of (001) planes increases by one from (a) to (c) in both electrodes 
increasing the size of the pyramids and the atomic detail
in the electrode bulk. The distance between pyramid apex atoms is 4.8\AA}
\label{tip-chain-tip}
\end{figure*}

\begin{figure*}
\vspace{-0.8cm}
\centerline {\epsfxsize=3cm \epsfysize=3cm 
\epsfbox{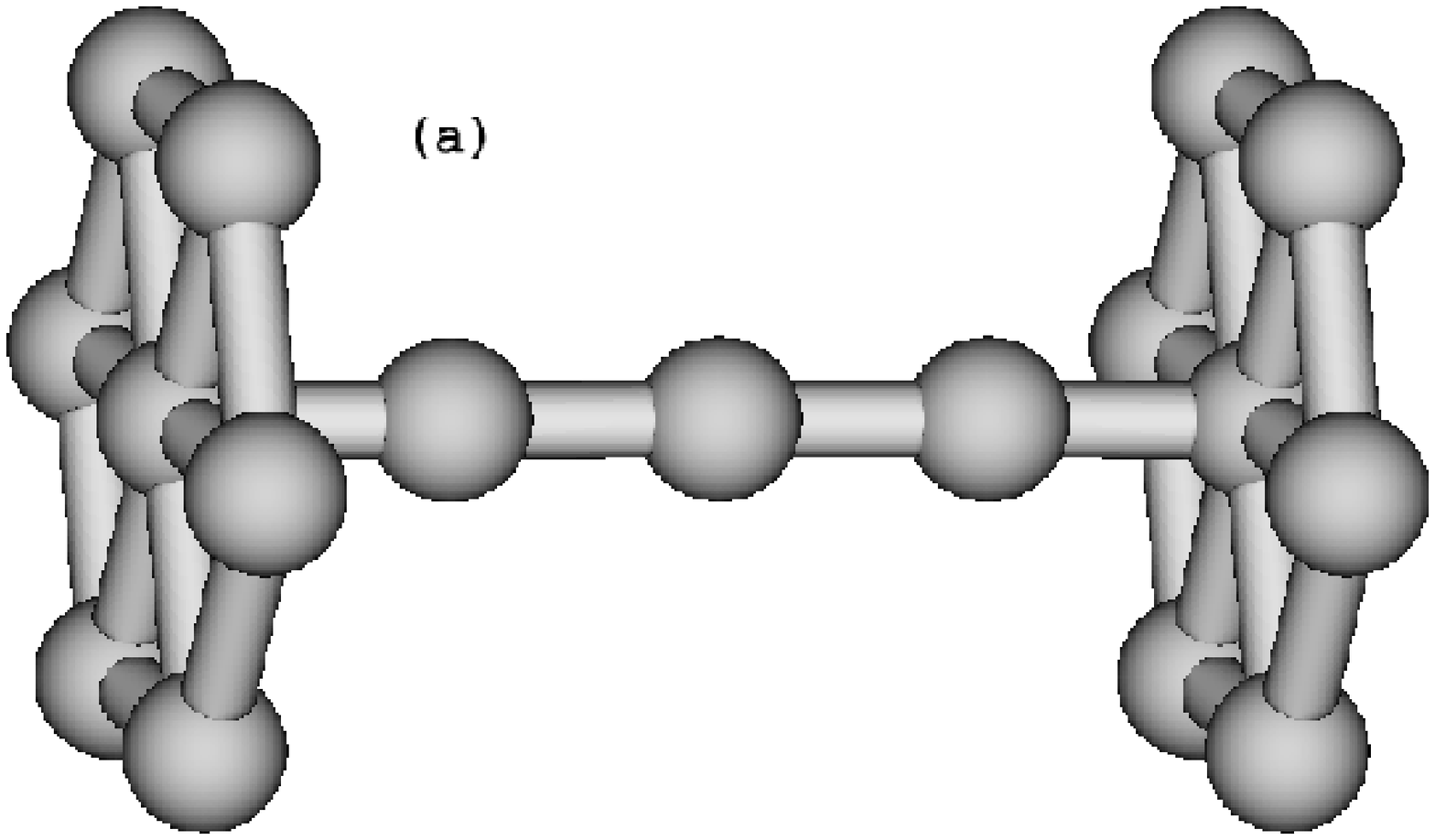}}
\vspace{-0.3cm}
\centerline {\epsfxsize=3cm \epsfysize=3cm
 \epsfbox{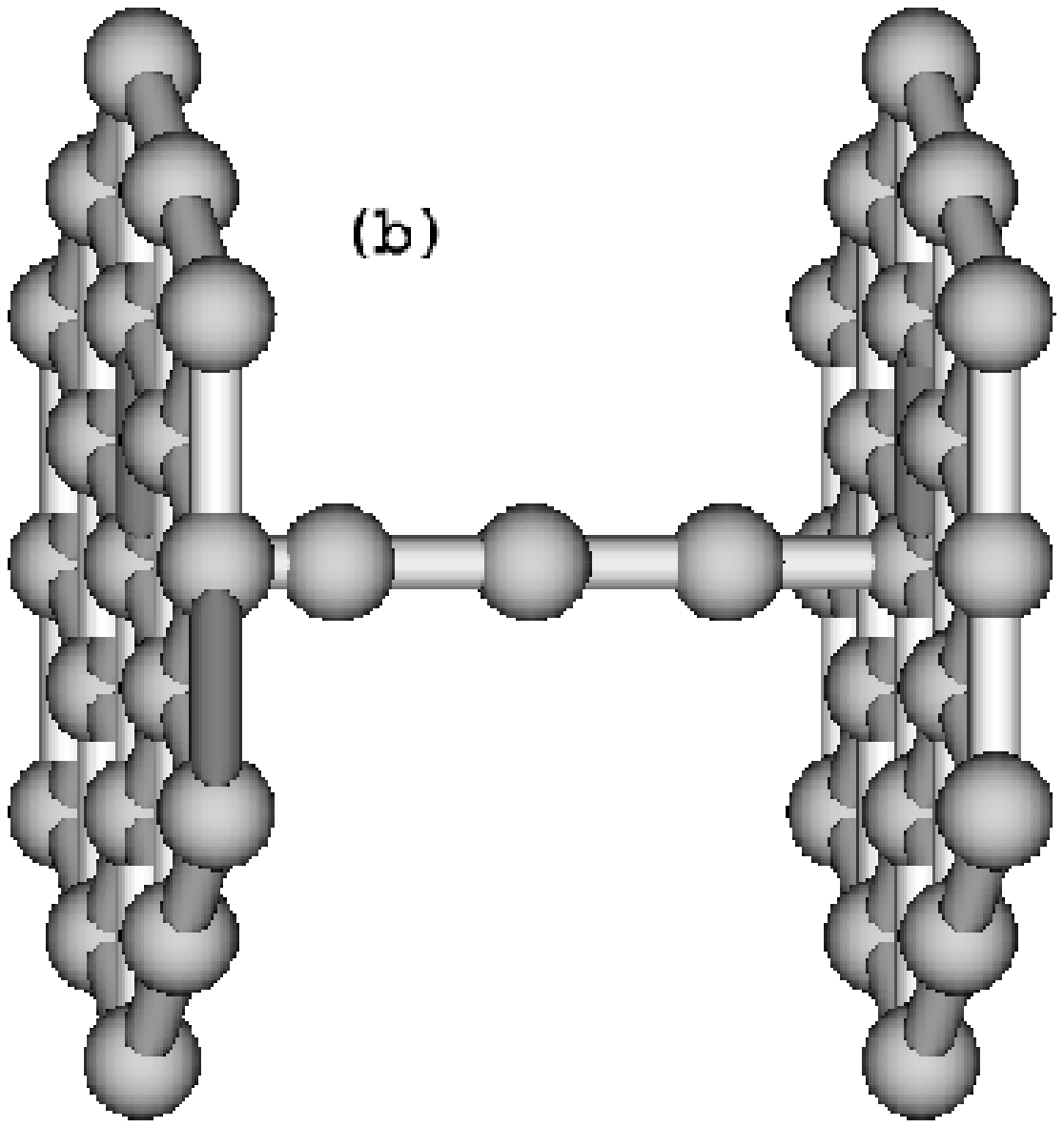}}
\vspace{-0.2cm}
\centerline {\epsfxsize=3cm \epsfysize=3cm
 \epsfbox{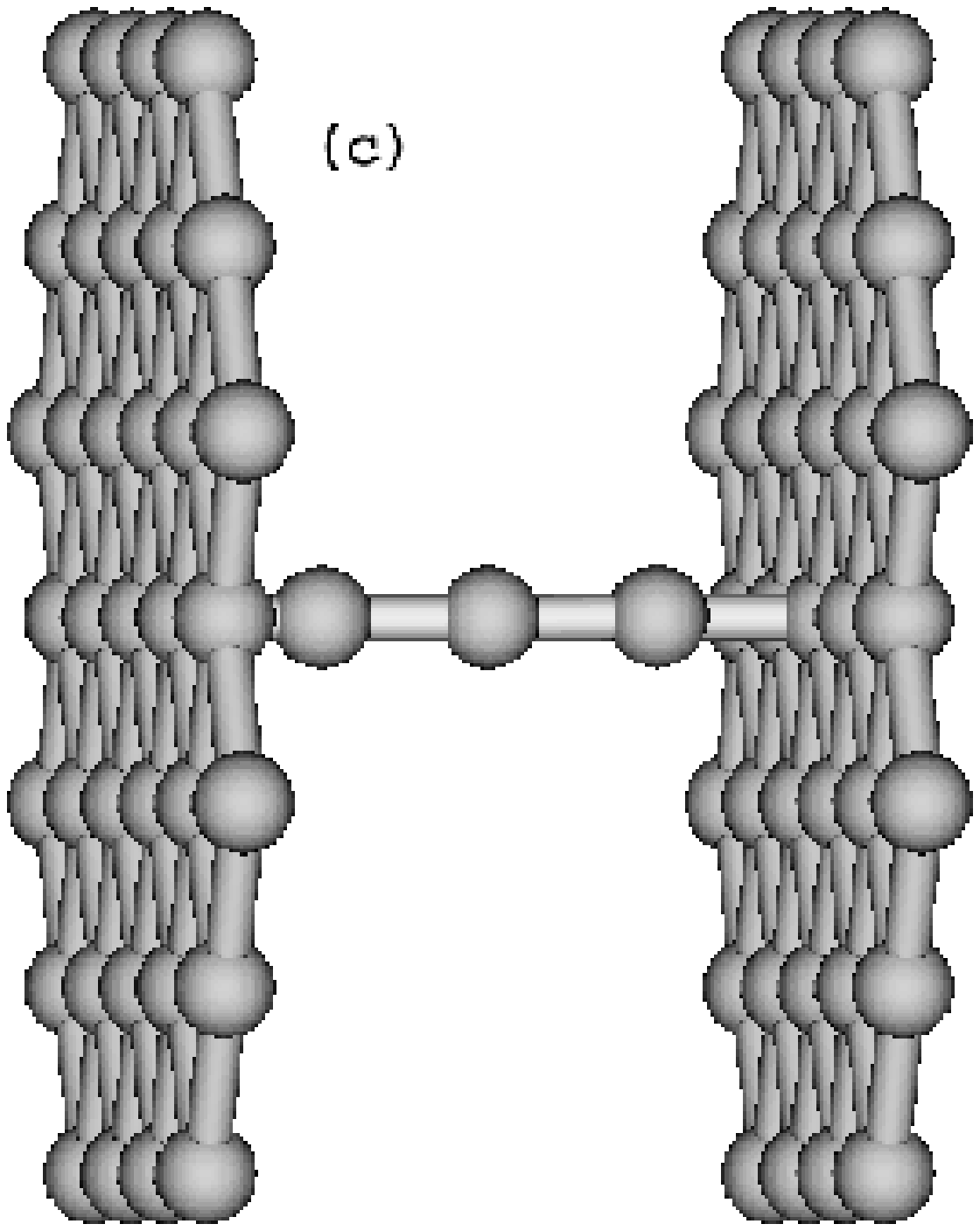}}
\vspace{-0.1cm}
\caption{Atomic structure of the second 
atom-chain constriction model considered. 
The number of atoms in the (111) electrode surface
increases from (a) to (c) in both electrodes 
increasing the atomic detail of the surface. The distance between planes
is 9.1\AA \ and the positions
of the atoms in the chain have been optimized.}
\label{111-chain-111}
\end{figure*}

A complete theoretical
study of electrical transport in metallic atomic contacts requires
a realistic modelization of the formation 
process of these nanocontacts. Some structural 
studies  using molecular dynamics\cite{An01} for Al and
{\em ab-initio} relaxations for Al\cite{SP99} and Na\cite{NBHJ99}
have been reported.  This is, however, a problem beyond
the scope of this work. Here we consider
archetypical atomic structure models that are likely
to appear in the last stages of the formation process of  atomic contacts
before the break-up:  Single-atom contacts and atomic wires.
More specifically,
our first structure consists of two opposite pyramids grown in the (001) 
direction and ``glued''
by a single atom [see Fig.\ \ref{tip-atom-tip}]. Single-atom contacts have been
studied in the past with
modified tight-binding models\cite{Cu98}. We find to our surprise 
that, even in this simple case, our {\em ab-initio} 
results are qualitatively different from those obtained with these 
models, particularly for Al.  
Our second structure is similar to the previous one, but with 
a chain of three atoms instead of a single atom
[see Fig.\ \ref{tip-chain-tip}]. Finally we have studied the same
chain in between two (111) surfaces with the chain placed on top of 
a surface atom [see Fig.\ \ref{111-chain-111}]. 
A similar geometry has been recently studied 
with {\em ab-initio} techniques for Au and Al. Our results
agree with what has been reported for Al using a jellium model for the 
electrodes\cite{KAT01}, 
but not entirely with what has been obtained for Au\cite{BT01,MW01}.

\subsubsection{Aluminum}
\begin{figure}
\vspace{-0.2cm}
\centerline {\epsfxsize=9cm \epsfysize=9cm \epsfbox{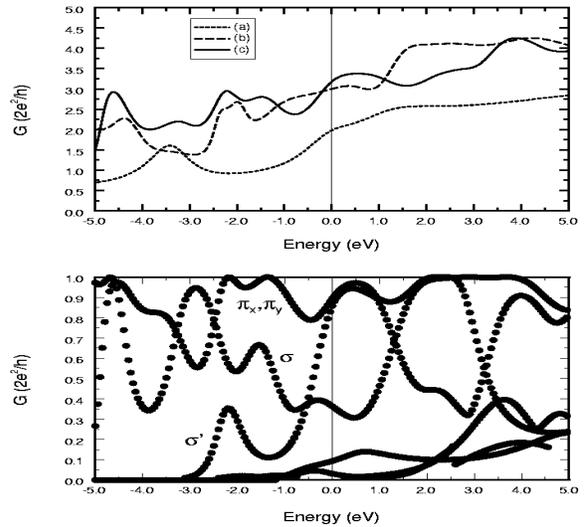}}
\vspace{-0.3cm}
\caption{Top: Conductance versus energy (Fermi energy set to zero) 
of the single-atom Al contact seen in Fig.\ \ref{tip-atom-tip} for the 
three cases shown there. Bottom: Individual contribution of the different
active conduction channels for (c). 
The labels indicate the orbital nature of the channels. The primed
label is associated with 2nd-nearest-neighbors hoppings.}
\label{altipatom}
\end{figure}

Figure \ref{altipatom} (top panel) shows the conductance vs. energy
for a single-atom Al contact. 
We have considered a $3s3p$ basis set and bulk interatomic distances. 
Curves (a), (b), and (c)
correspond to clusters  (a), (b), and (c) in  Fig. \ref{tip-atom-tip}, 
respectively.  In all the clusters the contact between electrodes occurs
through a single atom, but the number of (001) planes explicitly
included in the DF calculation for
each pyramid increases from (a) to (c) (remember that Bethe lattices
are always attached to the outer planes as in Fig.\ref{bethe}).
It is impossible 
to know the actual atomic structure of the metallic contact in detail unless
relaxation calculations are performed, but we do not expect the detailed 
geometry away from the neck to be important. In fact,
as Fig. \ref{altipatom} shows, the conductance does not change
significantly from (b) to (c), apart from minor changes in
the fine structure. This is a clear indication that, 
to a good extent, the conductance is determined by the 
atomic structure in the narrowest region of the neck. However, from 
our results we see that, at least, the nine central atoms must be explicitly
considered in this example. This is in contrast to the conclusions  drawn in 
Ref. \onlinecite{Cu98} by Cuevas 
{\em et al.} using a modified tight-binding model. 
Furthermore, the value of $G$  around the Fermi level is $\approx 3$  which is 
remarkably different from the value they obtained.  This discrepancy is due to 
a combination of facts. First, the hopping parameters that reproduce bulk 
properties in tight-binding models are not adequate for atoms with low 
coordination numbers, being these
typically smaller than the {\em ab-initio} ones.
Second, there are non-negligible contributions from next-to-near-neighbor 
hoppings in Al.  This is clearly seen in  the bottom panel of 
Fig.\ \ref{altipatom} where the contributions to the total current
of the main individual channels for the cluster (c) are depicted. 
As shown in the figure, a channel associated with 2nd-nearest-neighbors
hoppings (labelled $\sigma '$) can give a contribution of almost
one conductance quantum  $G_0$ ($G_0= 2e^2/h$) at the Fermi level. 
The main contribution at $\epsilon_F$ comes  from  two degenerate
$p_x,p_y$-like channels ($\pi_x,\pi_y$) that account for almost two 
conductance quanta $2G_0$. (The $z$-axis has been chosen along the main
symmetry axis of the cluster).
In addition there are two channels that have $sp_z$ character ($\sigma$) with 
non-negligible contributions at the Fermi energy that add approximately
0.3$G_0$ and 0.1$G_0$  to the total conductance, 
respectively. None of this seems
consistent with the tight-binding results for a similar geometry\cite{Cu98}.
\begin{figure}[tb]
\vspace{-0.2cm}
\centerline {\epsfxsize=9cm \epsfysize=9cm \epsfbox{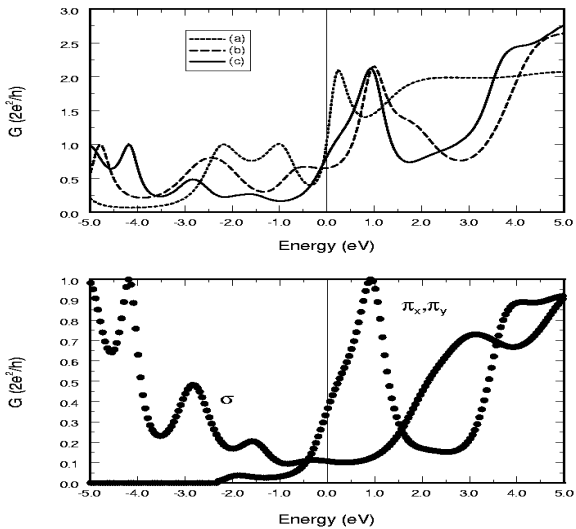}}
\vspace{-0.3cm}
\caption{Same as in Fig. \ref{altipatom},
but with three Al atoms forming a linear chain in between the electrodes
(see Fig.\ \ref{tip-chain-tip}).}
\label{altipchain}
\end{figure}    
The conductance of the three-atom chain 
shows a different behavior from the one in the previous example
and presents the same features
for the two electrode models considered (see top panels in
Figs.\ \ref{altipchain} and \ref{al111chain}). For the first electrode model
we have chosen a separation of 4.8\AA \ between apex atoms of the pyramids.
The conductance changes appreciably 
from (a) to (b) in Fig.\ \ref{altipchain}, but it does not so from (b) to (c).
(This dependence is similar to that in
 the single-atom contact). In the second case we have chosen a separation 
of 9.1\AA \ between planes and we have performed an {\em ab-initio}
relaxation of the chain
atom positions (only the surface layer is included in the DF calculation where
the number of atoms increases from 7 to 35). 
Here the conductance does not depend too much on the number
of  surface atoms [see Figs.\ \ref{111-chain-111}(a), (b), and (c)]. 
In all the cases there are oscillations as a 
function of the energy which, as the bottom panels in 
Figs.\ \ref{altipchain} and \ref{al111chain} show, appear mostly
in the $\pi$ channels. This is reminiscent of the behavior of the transmission
in  a Fabry-Perot interferometer due to
scattering at the  interfaces. These results are 
similar to the ones reported in 
Ref.\ \onlinecite{KAT01} where a jellium model was used to 
describe the Al electrodes. 

In the three cases considered there is no trace of possible
conductance quantization.  A general trend that 
can be observed in our results for the chain is  that
the onset of the transmission through the $\pi$ channels 
occurs close to the Fermi energy\cite{KAT01}. 
This makes $G$ strongly dependent on small variations in the 
positions of the atoms in the chain and on the atomic 
structure of the electrodes close to the chain.  
This might explain why the experimental conductance histograms
for Al are much more irregular than those for Au (see below) and other 
metals\cite{Ag93,Sc98}. However, there are many open questions regarding
the details of the conductance steps for Al that
illustrate the necessity of performing both relaxation and
conductance calculations at the same time\cite{AP02}.

\begin{figure}[tb]
\vspace{-0.2cm}
\centerline {\epsfxsize=9cm \epsfysize=9cm \epsfbox{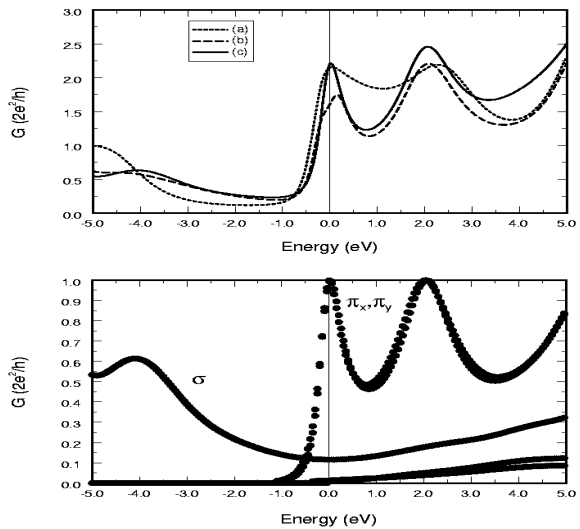}}
\vspace{-0.3cm}
\caption{Same as in Fig. \ref{altipchain},
but here the pyramids have been substituted by (111) surface planes. The
number of atoms included in the planes are 7(a), 19(b), and 35(c)
(see Fig.\ \ref{111-chain-111}).  }
\label{al111chain}
\end{figure} 

\subsubsection{Gold}
\begin{figure}[tb]
\vspace{-0.2cm}
\centerline {\epsfxsize=9cm \epsfysize=9cm \epsfbox{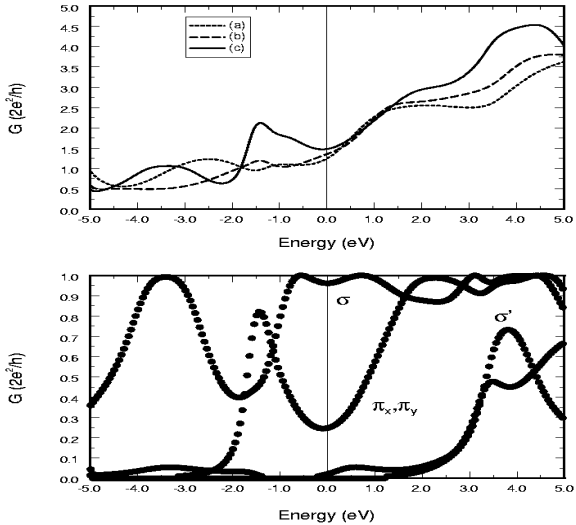}}
\vspace{-0.3cm}
\caption{Top: Conductance versus energy (Fermi energy set to zero) 
of the single-atom Au contact seen in Fig.\ \ref{tip-atom-tip} for the 
three cases shown there. Bottom: Individual contribution of the different
active conduction channels for (c). 
The labels indicate the orbital nature of them and $\sigma '$ indicates 
2nd-nearest-neighbors.}
\label{autipatom}
\end{figure}

We have repeated the conductance calculations for the same structures
considered above, but now consisting of Au atoms (we have used 
a here $5d6s6p$ basis set). In principle, only the electron of the $6s$ 
orbital is expected to contribute to the conductance at the Fermi energy which
should make the analysis of conductance simpler.
As Fig.\ \ref{autipatom} shows, the conductance around
the Fermi energy for the single-atom contact
varies little from (a) to (c).  We note again that, contrary to the 
tight-binding predictions\cite{Cu98}, the conductance at the Fermi energy
surpasses $G_0$ in all the curves. As concerns the
contribution of the individual channels we note that the
major contribution (almost a conductance quantum) 
has $sp_zd_0$ character ($\sigma$). Two degenerate
channels ($\pi_x,\pi_y$) of $p_xd_1$ and $p_yd_{-1}$
character (mainly $d$) give around 0.25$G_0$ 
quanta each. This should be expected since the $d$ orbitals contribute
significantly to the density of states in bulk atoms and the number
of near neighbors (eight) of the central atom
in this cluster is almost the bulk coordination number of an fcc 
structure (twelve).  Of the two channels that give a significant contribution
at rather high energies (above 2.0 eV) one has $sp_zd_0$ 
symmetry ($\sigma$) and the other corresponds to 2nd-nearest-neighbors.

It is interesting to compare the single-atom contact results for Al and Au. 
As remarked above, whereas in the case of Al there was a very
important contribution to the current at the Fermi level coming from
2nd-nearest-neighbors hoppings, in Au this was only appreciable
at high energies. This cannot be understood in terms of the 
respective atomic radii which are very similar (1.43 \AA
and 1.45 \AA for Al and Au, respectively\cite{FK76}), but
rather it is due to the character of the wavefunctions at
the Fermi level in each case. Namely, while in Al the density
of states at $E_F$ mainly comes from $p$ orbitals that
are rather extended, in Au the wavefunction at that energy mainly has 
an $s$ character and is more localised.  
\begin{figure}
\vspace{-0.2cm}
\centerline {\epsfxsize=9cm \epsfysize=9cm \epsfbox{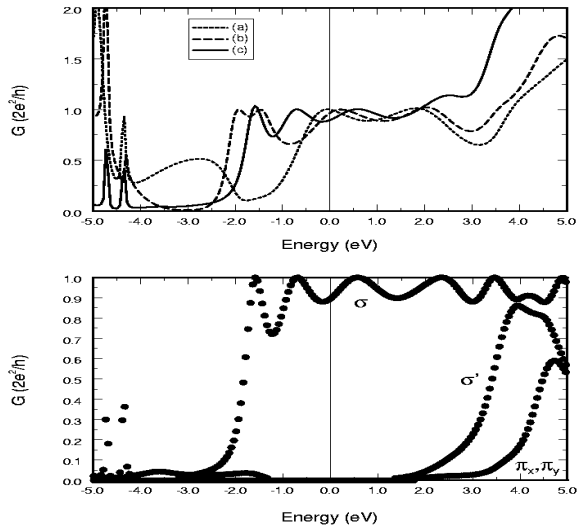}}
\vspace{-0.3cm}
\caption{Same as in Fig. \ref{autipatom},
but with three Au atoms forming a linear chain in between the electrodes
(see Fig.\ \ref{tip-chain-tip}).}
\label{autipchain}
\end{figure}    
On the other hand, the conductance around the Fermi energy for the
three-atom 
Au chain shows clearly an upper limit of $G_0$  (see top panels in
Figs.\ \ref{autipchain} and \ref{au111chain}),
and it does not change qualitatively with the cluster size. Nevertheless, the 
exact value at the Fermi energy is elusive, changing by as much as 20\% 
from cluster to cluster.  We have not been able to
verify whether the conductance curves for larger clusters converge to a given
one, but all the curves present a characteristic behavior: Above the Fermi
energy the conductance is fairly constant while below oscillates and vanishes
right above the $d$ channel contribution.  The channel decomposition analysis
is quite simple: A single $sp_zd_0$-like channel ($\sigma$) contributes around 
the Fermi energy.
Nevertheless, it is still difficult to explain from our results 
the robustness of the quantization observed in the experiments\cite{Ag93,Sc98} 
which does not deviates
from $G_0$ by more than a small percentage over a large range of stretching
force.  Large-scale structural studies along with conductance calculations
are also desirable here in order to make a precise quantitative
comparison with experiments. Recent {\em ab-initio}
works\cite{BT01,MW01} for Au nanocontacts have partially 
addressed this problem. However, as already pointed out, the electrode model  
considered there seems to introduce serious difficulties  in the 
interpretation of their conductance results.
\begin{figure}
\vspace{-0.2cm}
\centerline {\epsfxsize=9cm \epsfysize=9cm \epsfbox{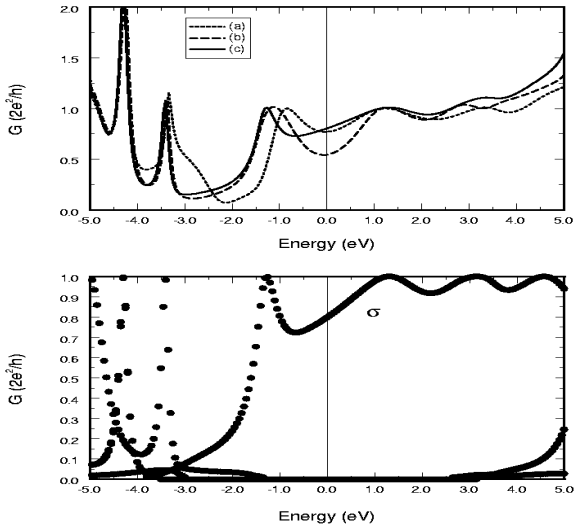}}
\vspace{-0.3cm}
\caption{Same as in Fig. \ref{autipchain},
but here the pyramids have been substituted by (111) surface planes. The
number of atoms included in the planes are 7(a), 19(b), and 35(c)
(see Fig.\ \ref{111-chain-111}).  }
\label{au111chain}
\end{figure} 

\subsection{Carbon chains}
The conductance of C-atom chains attached to Al electrodes has been calculated
from first principles in previous works by Lang and Avouris\cite{LA98,LA00}. 
In their calculations semi-infinite jellium models were used
to describe the metal electrodes and a pseudopotential for the C cores. The 
self-consistent density functional procedure they used (see Refs. 
\cite{LA98,LA00} for details) predicted an oscillatory behavior in the 
conductance at the Fermi level with maxima (minima) in those chains that
had an odd (even) number of electrons. This oscillatory behavior differs from
that corresponding to a closed shell electronic structure and an $sp_z$ 
hybridization of linear C chains. In this situation each C atom 
added to the chain 
provides two $sp$ orbitals that contribute to the $\sigma$ molecular 
orbitals (MOs) and two $p$ orbitals that contribute to the corresponding 
$\pi$ MOs. With a closed-shell electronic structure and an {\em even} number of 
C atoms there is a partial filled $\pi$ shell, and the chain could be 
considered to be intrinsically a conductor. On the other hand, if the chain has 
an odd number of C atoms, the $\pi$ and $\sigma$ shells are completely 
filled and we would have a semiconductor. Thus, one would expect that chains
with an {\em even} number of C atoms would provide higher
conductances than chains with an {\em odd} number of C atoms\cite{LA98}. 
However, either if we consider an open shell electronic structure for the
C chain, or if the edge C atoms are bonded to a metal surface, the 
MOs can be filled in a different way, which may lead to an {\em inversion} 
of the aforementioned trend: that is, chains with an {\em odd} number of C 
atoms having a partially filled $\pi$ shell while chains with an {\em even} 
number of C atoms having this shell completely filled. The first scenario,
i.e., an open-shell electronic structure, is actually the ground-state for
isolated linear C chains, while the second one explains the findings of
Lang and Avouris for the conductance of C chains attached to Al
electrodes\cite{LA98,LA00}. In this second situation  the details of the
bonding region between the C and metal atoms is of primary importance.
This involves knowing the geometry of the contact, the C-metal surface
distance, and the electronic structure of the metal under interest. This,
however, cannot be easily deduced by using a jellium model to
describe the leads.

\begin{figure}
\centerline {\epsfxsize=4cm \epsfysize=4cm \epsfbox{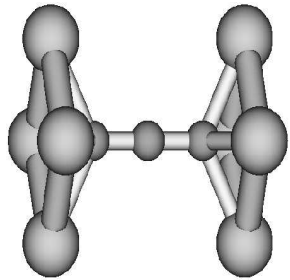}}
\vspace{0.6cm}
\caption{Atomic structure of a C$_3$ chain attached to  hollow sites in
between (001) fcc Al surfaces.}
\label{c3-chain}
\end{figure} 

For this reason, we have applied the method described in the previous sections 
to these kind of 
systems. More precisely, we have calculated the conductance for C chains 
starting from 3 C atoms attached to Al and Au (001) fcc electrodes. 
The distance between C atoms was fixed to that used in Refs. 
\onlinecite{LA98,LA00}, i.e., 
an equal spacing of 2.5 a.u. between C atoms. The description of the
electrode surface was reduced to four metal atoms describing a hollow site
(see Fig. \ref{c3-chain}) in the center of which the C chain was attached. 
The importance of the C-Al surface distance was analyzed by making two sets of
calculations for each C chain. In the first set this distance was kept
fixed at a reference value of 2\AA, while in the second one the distance
between the C chain and the metal electrode was that providing the minimum
energy of the corresponding cluster. The results are commented next.

\begin{figure}
\centerline {\epsfxsize=5cm \epsfysize=4cm
 \epsfbox{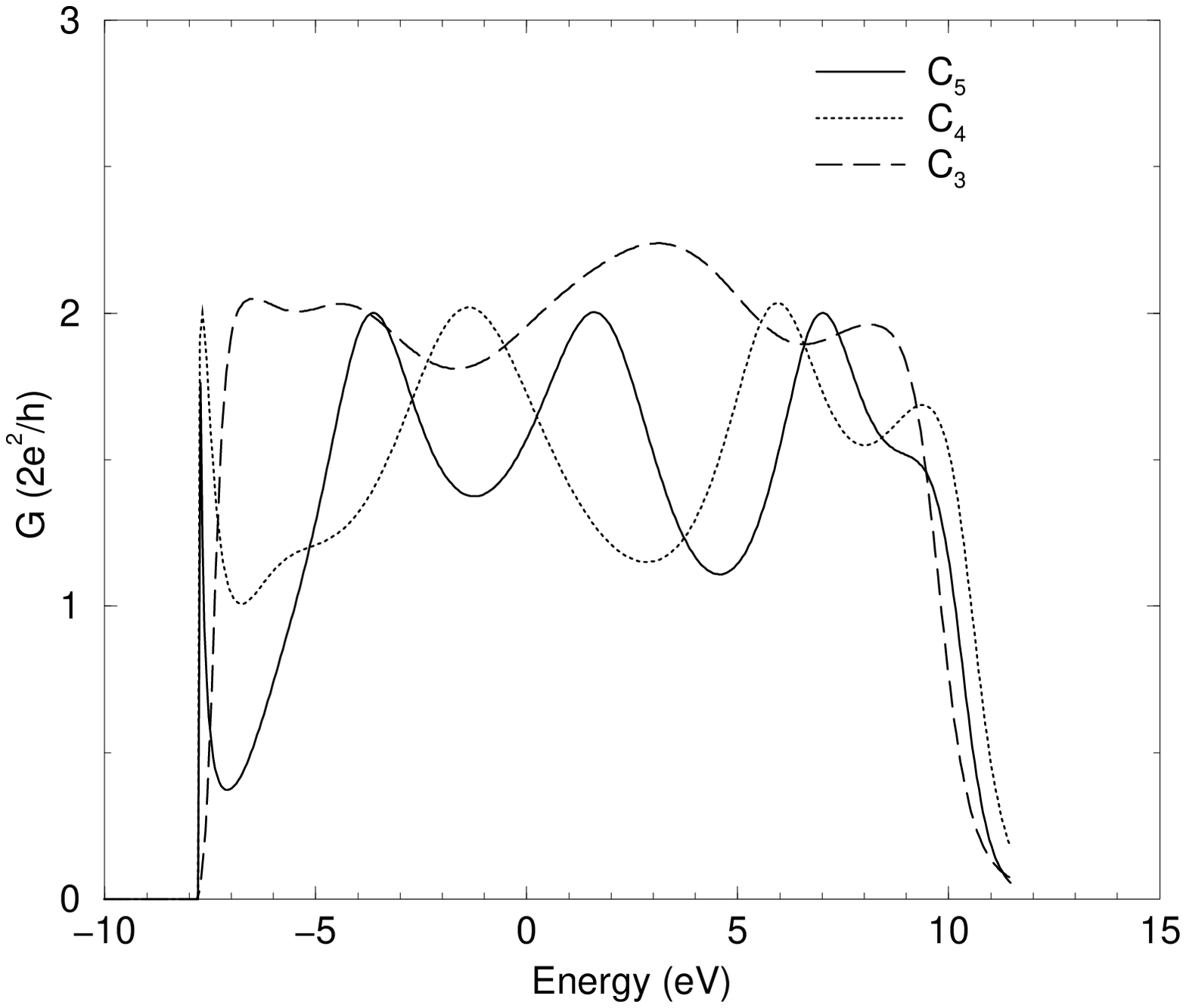}}
\centerline {\epsfxsize=5cm \epsfysize=4cm
 \epsfbox{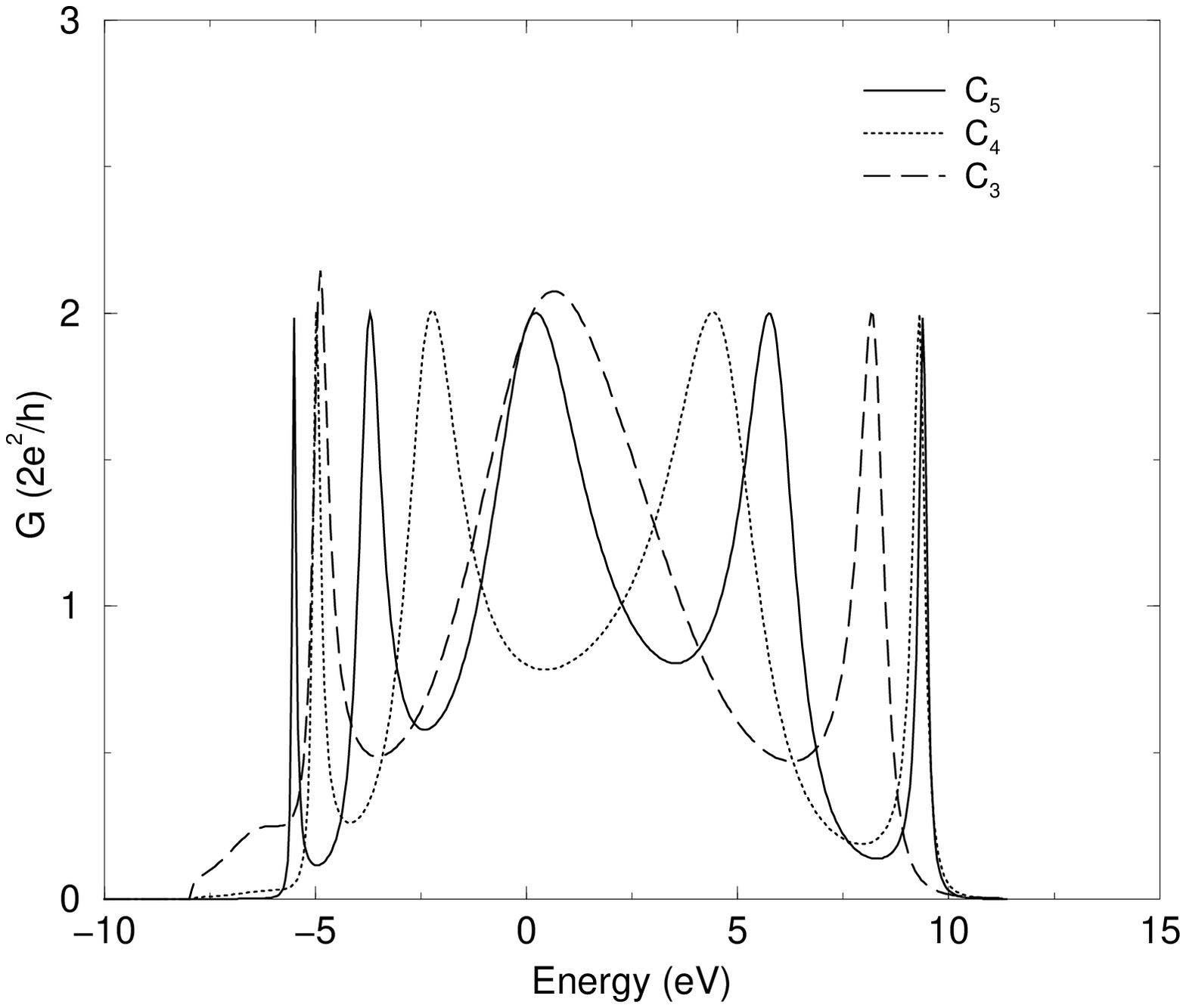}}
\caption{Top: conductance versus energy (Fermi energy here set to zero)
of C$_{n}$ linear chains ($n=3-5$) attached to Al at equilibrium C-Al distance
(1\AA).
Bottom: the same at a distance C-Al of 2\AA.}
\label{GE-C-Al}
\end{figure}    

\subsubsection{Al electrodes}
The discussion about the conductor or semiconductor character of C chains
given in the preceding section can be easily extended to the conductance of 
these chains once they have been contacted to semi-infinite electrodes. 
Note that
the oscillatory character of the conductance at the Fermi level with the number 
of C atoms $n$ in the chain found in these systems can be understood as 
a consequence of the oscillatory nature of the conductance with respect to the 
energy (reflected in Fig. \ref{GE-C-Al}). The oscillations in $G(E)$ 
come from {\em two} 
degenerate $\pi$ channels, thus explaining the maximum value of 2$G_0$ 
(the only exceptions being the C$_{3}$ and C$_{4}$ chains where additional 
$\sigma$ channels appear around the Fermi energy). The number of peaks is 
given by the number of $\pi$ orbitals in the chain which, 
in turn, depends on the number of C 
atoms present in the chain. The positioning of the Fermi level near the 
center of these peaks (partially filled $\pi$ shell) or in between peaks
(completely filled $\pi$ shell) would determine the oscillatory trend with
$n$. On the other hand, both the exact
position of the conductance peaks with respect 
to the Fermi level and their average size are governed by the distance between 
the edge of the C chain and the Al electrode.

\begin{figure}
\centerline {\epsfxsize=5cm \epsfysize=4cm
 \epsfbox{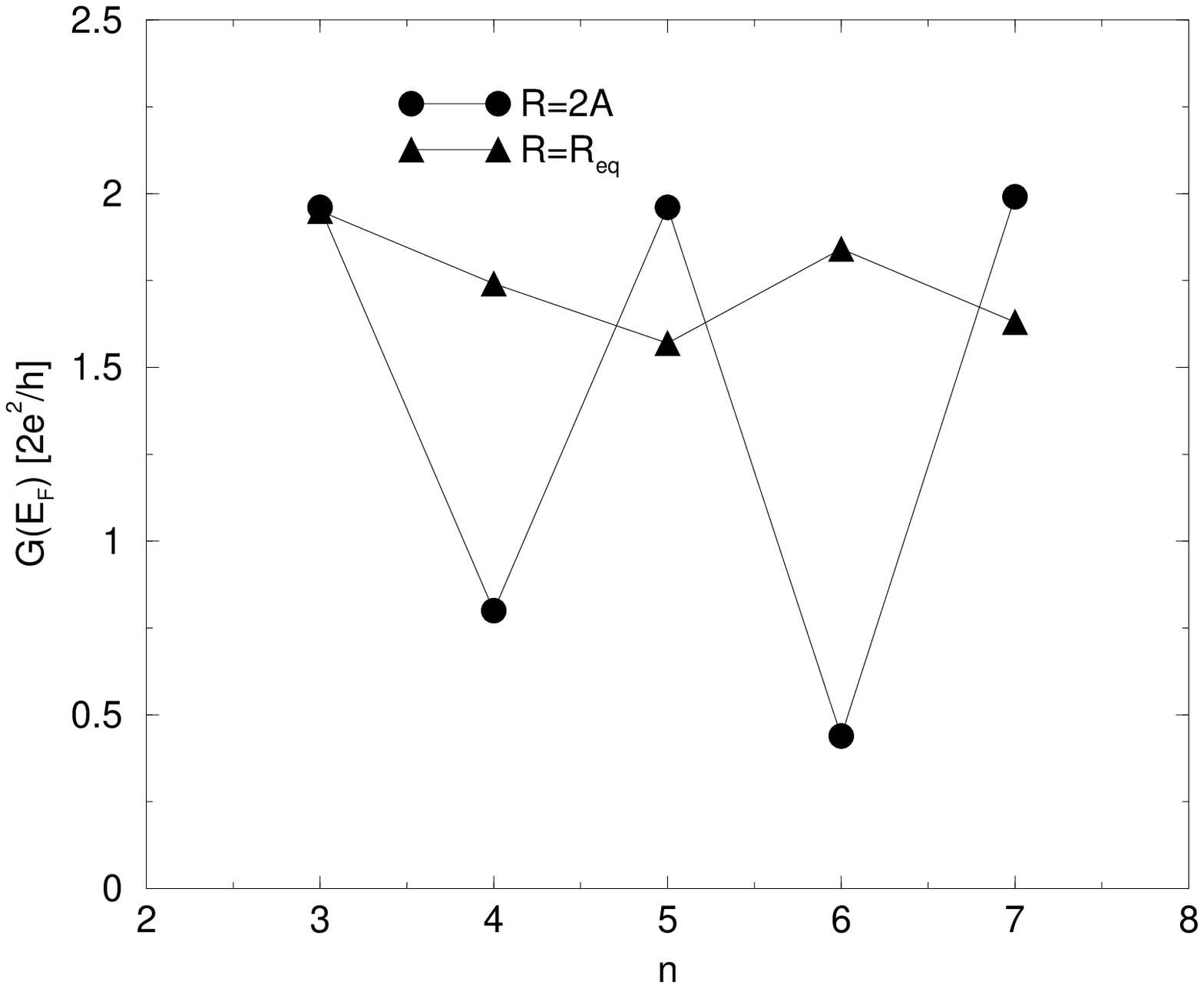}}
\centerline {\epsfxsize=5cm \epsfysize=4cm
 \epsfbox{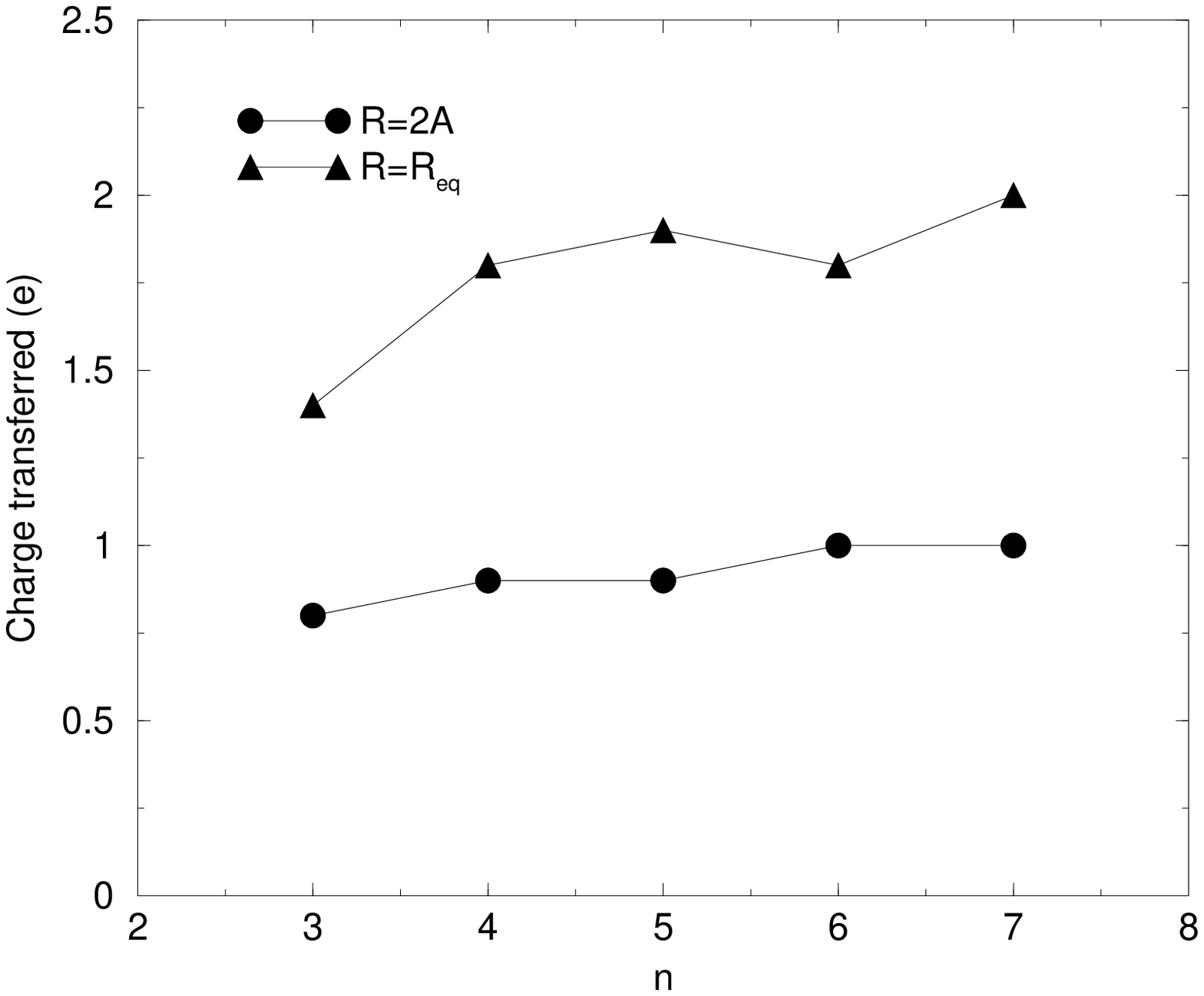}}
\caption{Top: conductance at the Fermi level of C$_{n}$ linear chains 
($n=3-7$) attached to Al electrodes. Bottom: charge transferred to the 
C chain.}
\label{GEFq-C-Al}
\end{figure}    

\begin{figure}
\centerline {\epsfxsize=5cm \epsfysize=4cm
 \epsfbox{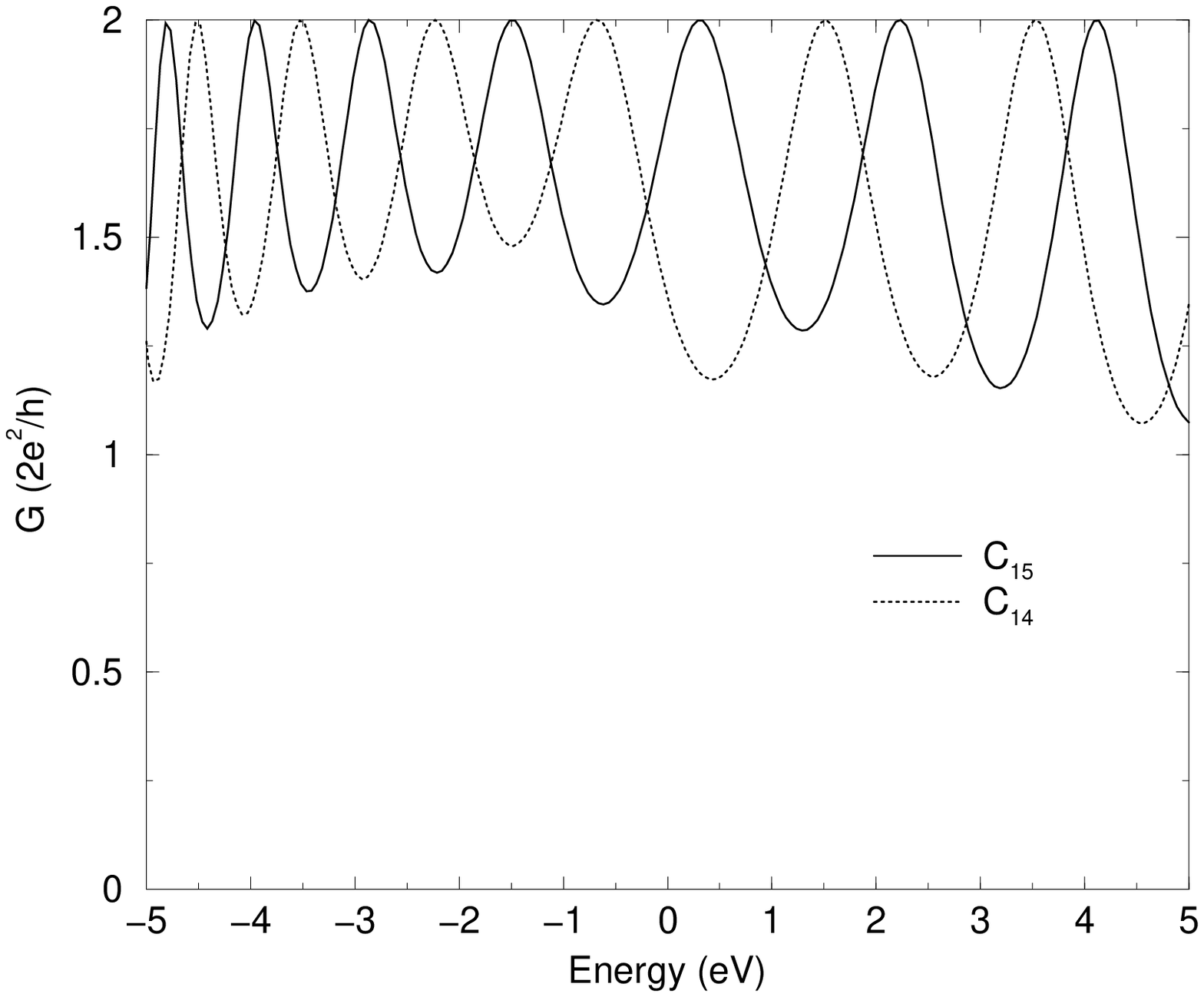}}
\centerline {\epsfxsize=5cm \epsfysize=4cm
 \epsfbox{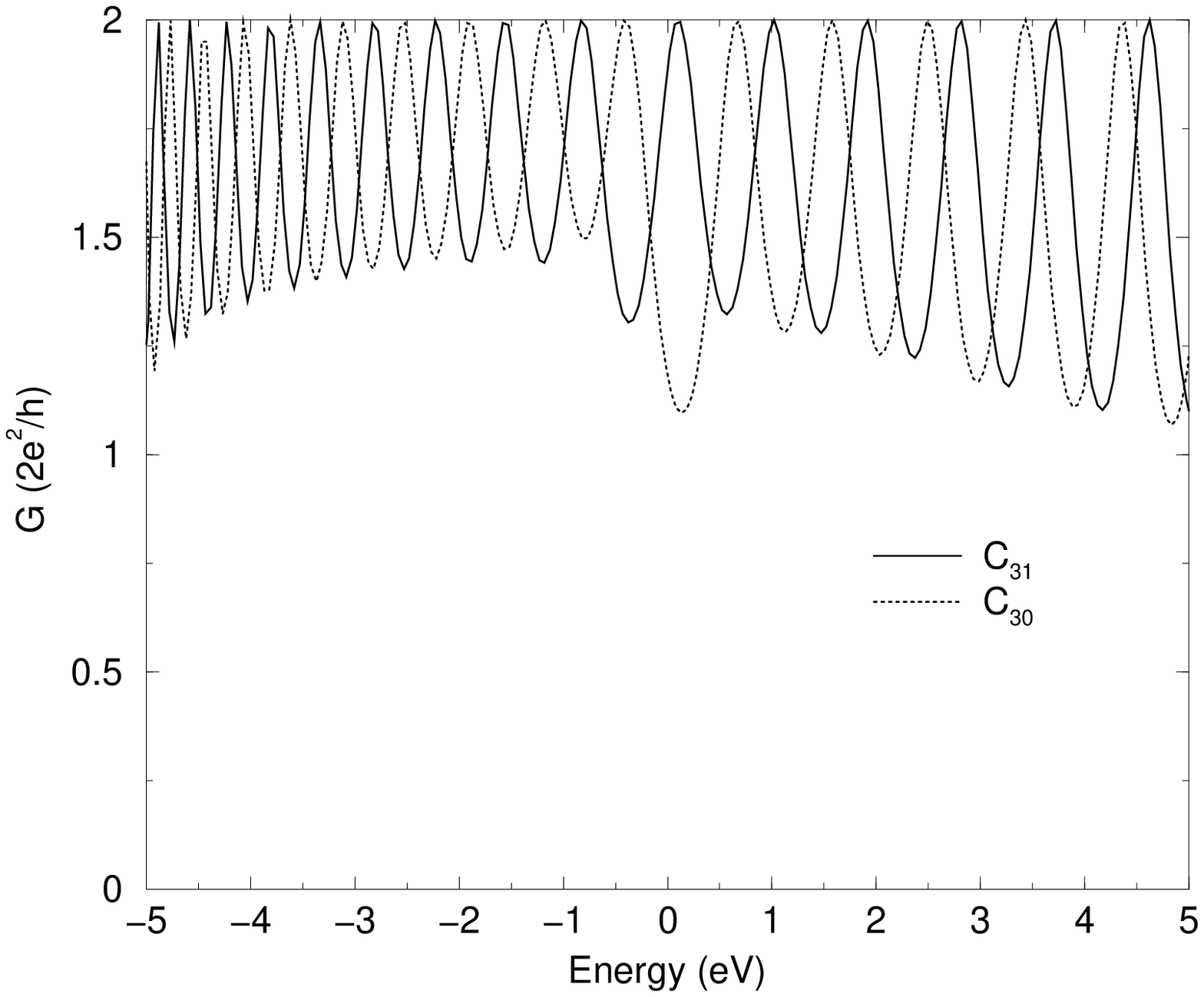}}
\caption{Top: conductance vs. energy (Fermi energy here set to zero) of 
C$_{n}$ linear chains 
($n=14,15$) attached to Al electrodes. Bottom: the same for $n=30,31$.}
\label{GE-C-Al-large}
\end{figure}    

After inspection of Fig. \ref{GE-C-Al}, where we plot the $G(E)$ for 
the two sets of calculations mentioned at the beginning of this section, 
it is evident that at
a distance of 2\AA between the edge C atom and the Al metal 
surface we end up with the same situation found by Lang and Avouris 
\cite{LA98,LA00}, namely, maxima (minima) located at $n$ odd (even) (see 
Fig. \ref{GEFq-C-Al}).
Nevertheless, at the equilibrium distance, where the C-Al surface distance
is smaller (1\AA), and the transferred charge from the metal to the C chain is
larger, the situation has changed and now the maxima are located
on chains with an even number of C atoms (the only exception being again
the C$_3$ chain) as Fig. \ref{GEFq-C-Al} shows. 
However, one would expect that the 
influence of the bonding region, a local effect, would become less important as 
the size of the C chain increases. This is what is actually found in the 
conductance of larger number of atoms, such as those presented in Fig. 
\ref{GE-C-Al-large},
where we plot $G(E)$ for $n$=14,15 and for $n$=30,31 and a distance
C-Al surface of 1\AA. In both cases the chains with $n$ odd 
give larger values for $G(\epsilon_F)$ than those with $n$ even. 
The same behavior
is found for the same chains attached to Al electrodes at a distance of
2\AA,  but has not been included in the figures for simplicity. It is 
worth noting that, strictly speaking at zero temperature,
the peaks in $G(E)$ and, thus, the oscillations in $G(\epsilon_F)$
are expected to survive in the thermodynamic limit $n \rightarrow \infty$.
This effect will disappear
when the temperature becomes of the same magnitude as the peaks width, where  
$G(\epsilon_F)$ will be the average value for the odd- and even-$n$ chains.

\begin{figure}
\centerline {\epsfxsize=5cm \epsfysize=4cm
 \epsfbox{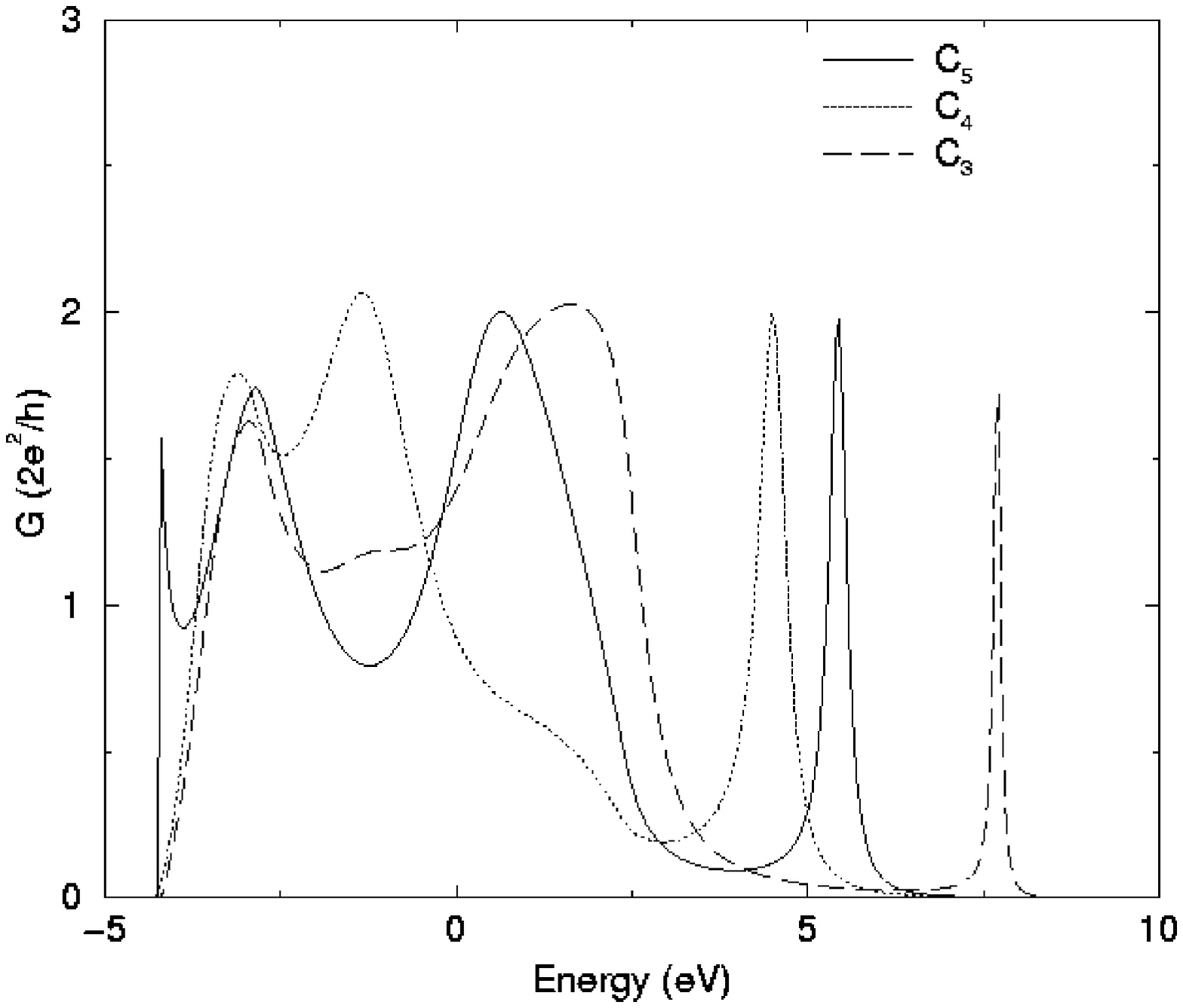}}
\centerline {\epsfxsize=5cm \epsfysize=4cm
 \epsfbox{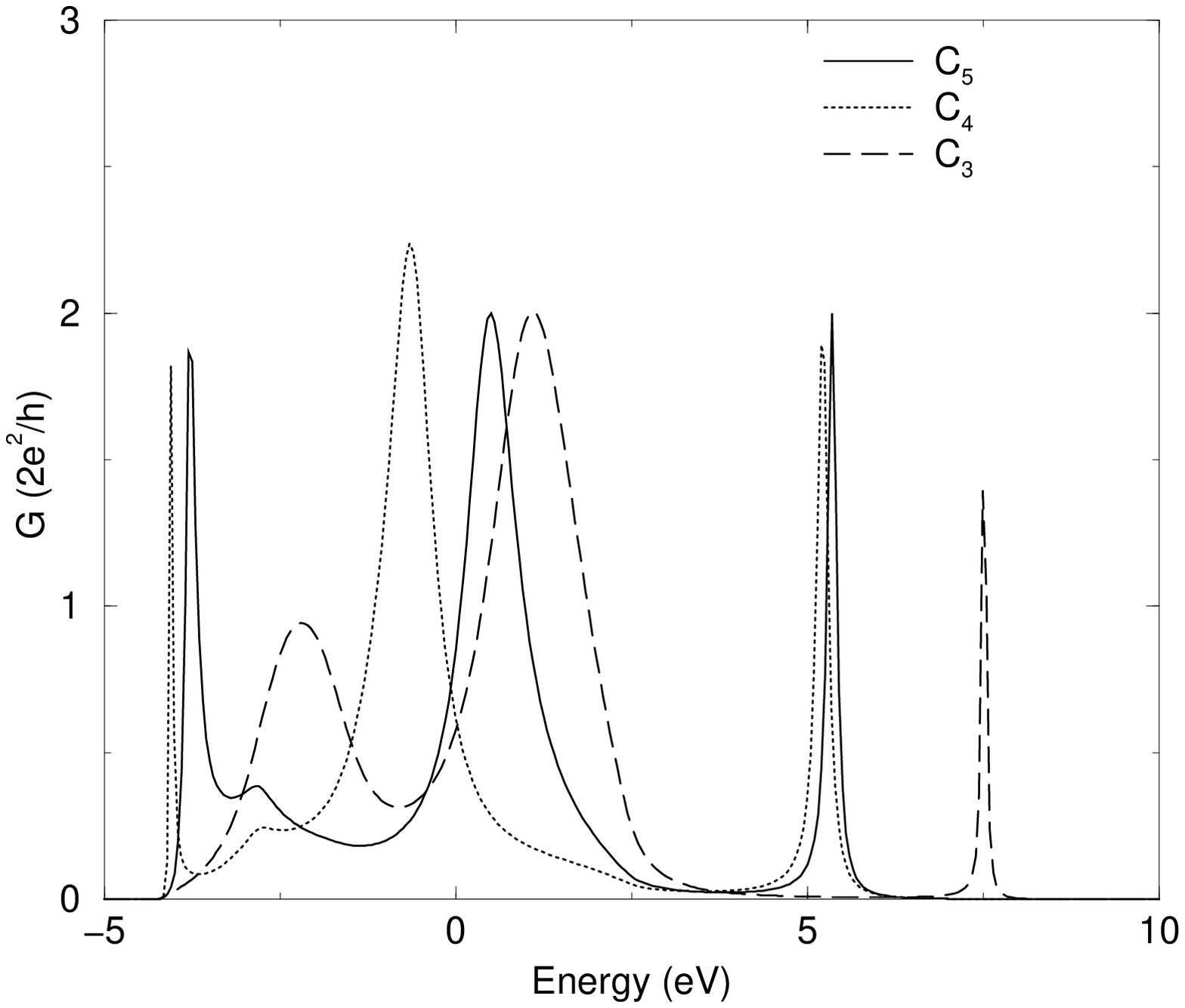}}
\caption{Top: conductance versus energy (Fermi energy here set to zero)
of C$_{n}$ linear chains ($n=3-5$) attached to Au at equilibrium C-Au
distance (1.5\AA). Bottom: the same at a distance C-Au of 2\AA.}
\label{GE-C-Au}
\end{figure}    

\subsubsection{Au electrodes}
\begin{figure}
\centerline {\epsfxsize=5cm \epsfysize=4cm
 \epsfbox{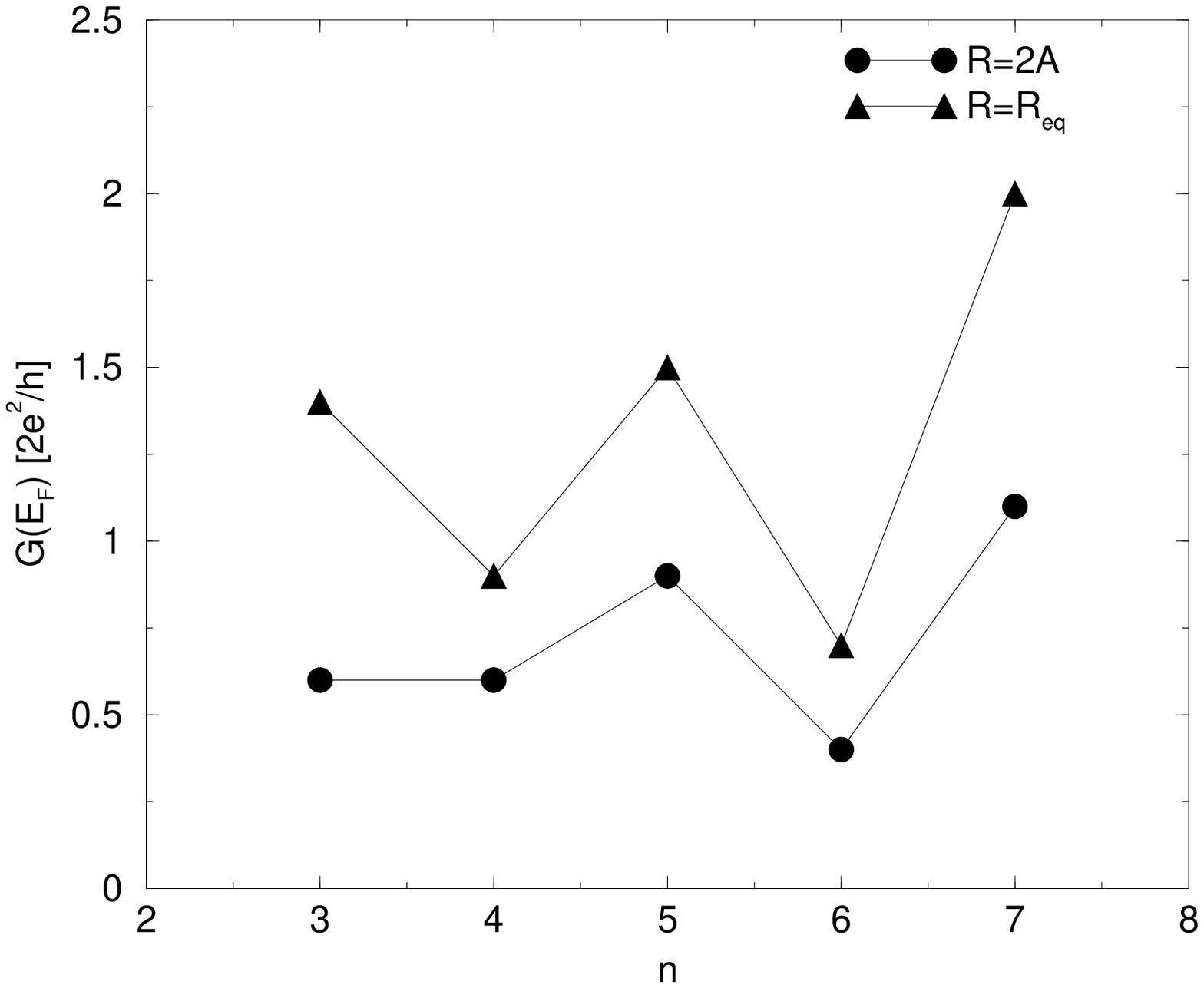}}
\centerline {\epsfxsize=5cm \epsfysize=4cm
 \epsfbox{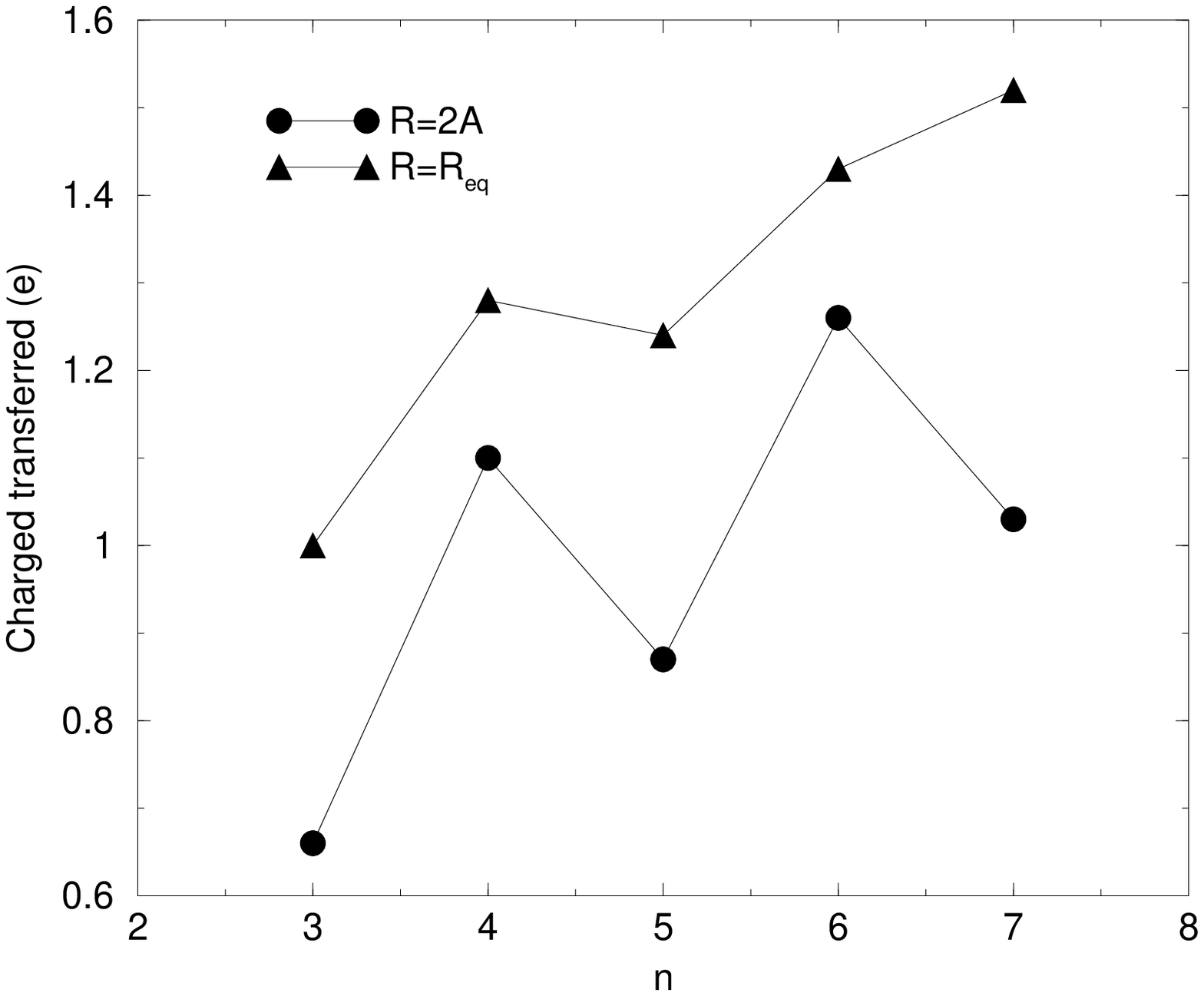}}
\caption{Top: conductance at the Fermi level of C$_{n}$ linear chains 
($n=3-7$) attached to Au electrodes. Bottom: charge transferred to the 
C chain.}
\label{GEFq-C-Au}
\end{figure}

Another key point in the conductance of nanocontacts is the nature of the
metal used in the electrodes which has been thoroughly discussed in previous
sections. This is specially the case when the chemical bond between the
molecule and the metal contact changes markedly as reflected in 
Fig. \ref{GE-C-Au} as compared to Fig. \ref{GE-C-Al}. There we show $G(E)$ 
for the same systems previously analyzed for Al but
with Au instead.  When we move from Al to Au the bonding between 
the C chain and 
the metal surface weakens. This reflects in a sharpening of the 
conductance peaks and in the amount of charge transferred from the metal
to the C chain (compare Figs. \ref{GEFq-C-Au} and \ref{GEFq-C-Al}). 
The equilibrium distance between 
the edge C atom and the Au surface is also typically 
larger in this case: 1.5\AA  \ for all the chains. As for Al, there also appear
two $\pi$ channels, except for the C$_{3}$ and C$_{4}$ chains.
However, the narrowing of the conductance peaks makes 
critical the alignment of the Fermi level. With respect to the oscillations of 
$G(\epsilon_F$), we observe the same tendency irrespective of the C-Au 
distance, that is: maxima located at $n$ odd. This is also a consequence of
the weaker bond between Au and C and the lesser amount of charge transferred
from the former to the latter.

\section{Concluding remarks}
In summary, we have developed a methodology to self-consistently
calculate {\em ab-initio} transport properties in atomic-scale systems
based upon the Gaussian98 code. This opens the doors to the use 
of the most standard quantum chemistry tool to the study of transport 
through molecules, an interdisciplinary  subject of increasing interest. 
We have chosen to study two systems that illustrate the capabilities  
of our approach: metallic constrictions of simple and noble metals 
and C chains with a reactive electrode (Al) and an ``inert'' (Au) electrode. In 
the first case, we have shown that assuming local charge neutrality, as commonly
done in semi-empirical methods, may lead to qualitatively incorrect results.
On the other hand, the system metal-C chain-metal illustrated how the chemistry
of the contact may determine electrical transport and, therefore, the 
incorrectness of ascribing one or another behavior to a given molecule leaving 
apart the specific electrode/molecule chemistry. One should speak, instead, of 
electrical transport characteristics of the whole system, 
electrode-molecule-electrode. 

\acknowledgements
Part of this work was supported by the 
Spanish CICYT under Grants Nos. 1FD97-1358 and PB96-0085,
and by the Generalitat Valenciana under Grants No. GV00-151-01
and GV00-095-2.  Discussions with L. Pastor-Abia,  J.
M. P\'erez-Jord\'a, J. C. Sancho, N. Agra\"{\i}t, P. Serena, A. Hasmy, 
D. S\'anchez-Portal, C. Untiedt, F. Guinea,
and J. J. S\'aenz are acknowledged. We are also grateful to 
D. A. Papaconstantopoulos for useful correspondence concerning the 
tight-binding parameters used in band structure calculations. 

\section{appendix}

In this appendix we discuss how selfenergies for  Bethe Lattices (BL) with 
no symmetry can be calculated. Symmetry can be broken due to either
the spatial atomic arrangement  or the orbitals on
the atoms that occupy each lattice site, or both. When no symmetry exists
the selfenergy in an arbitrary direction cannot be obtained by rotating
that for a given direction as done in Ref. \onlinecite{MV90}. Instead, the
following procedure has to be followed. The method is valid
for any basis set or lattice. Let ${\bf \tau_i}$ be the $N$
nearest-neighbor directions of the lattice we are interested in
and ${\hat V}_{\bf \tau_i}$ the interatomic interaction matrix in 
these directions.
The selfenergies associated to each direction have to be obtained
from the following set of $2N$ coupled selfconsistent equations,
\begin{mathletters}
\begin{equation}
{\hat \Sigma}_{\bf \tau_i}={\hat V}_{\bf \tau_i}
\left [ E{\hat {\bf 1}}-{\hat E}_0-
({\hat \Sigma}_{\bar T}-
{\hat \Sigma}_{\bar {\bf \tau_i}})\right ]^{-1}
{\hat V}_{\bf \tau_i}^{\dagger}
\end{equation}
\begin{equation}
{\hat \Sigma}_{\bar {\bf \tau_i}}={\hat V}_{\bar {\bf \tau_i}}
\left [ E{\hat {\bf 1}}-{\hat E}_0-({\hat \Sigma}_{T}-
{\hat \Sigma}_{\bf \tau_i})\right ]^{-1}
{\hat V}_{\bar {\bf \tau_i}}^{\dagger},
\end{equation}
\end{mathletters}
\noindent where $i=1,...,N$ and ${\bar {\bf \tau_i}}=-{\bf \tau_i}$. 
$E{\hat {\mathbf 1}}$ is the energy times the identity matrix, 
${\hat E}_0$ is a diagonal
matrix containing the orbital levels, ${\hat V}_{\bf \tau_i}$ 
is the interatomic
interaction in the ${\bf \tau_i}$ direction, and 
${\hat \Sigma}_{T}$
and ${\hat \Sigma}_{\bar T}$ are the sums of  the selfenergy 
matrices entering through all the Cayley tree branches attached to an 
atom and their inverses, respectively, {\it i.e.},   
\begin{mathletters}
\begin{equation}
{\hat \Sigma}_{T}=\sum_{i=1}^{N}{\hat \Sigma}_{\bf \tau_i}
\end{equation}
\begin{equation}
{\hat \Sigma}_{\bar T}=\sum_{i=1}^{N}{\hat \Sigma}_
{\bar{\bf \tau_i}}.
\end{equation}
\end{mathletters}
\noindent This set of $2N$ matricial equations has to be solved iteratively.
It is straightforward to check that, in cases of full symmetry,  
it reduces to the single equation discussed in \cite{MV90}.

\begin{figure}
\centerline {\epsfxsize=5cm \epsfysize=5cm
\epsfbox{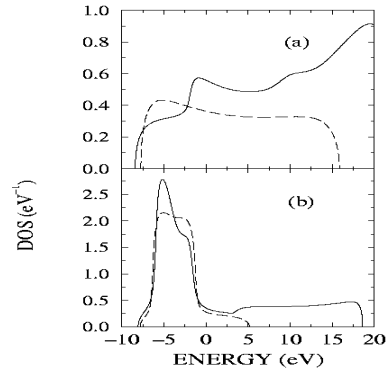}}
\caption{Density of states for the Bethe lattices of Al (a) and Au (b) 
obtained with the parameters of Table \ref{table}. Continuous lines correspond 
to results obtained with a $spd$ basis while broken lines to those 
obtained with either a $sp$ (Al) or a $sd$ (Au) basis. The Fermi level 
was set at zero energy.} 
\label{Bethe_Al_Au}
\end{figure}  
\begin{figure}
\centerline {\epsfxsize=5cm \epsfysize=5cm
\epsfbox{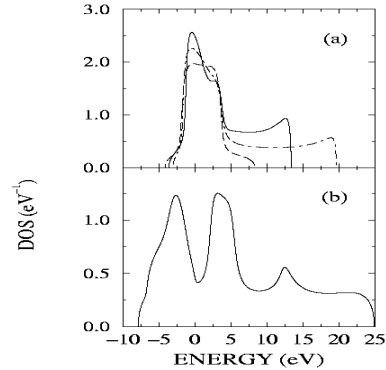}}
\caption{Density of states for the Bethe lattices of Ti (a) and W (b) 
obtained with the parameters of Table II. Continuous lines correspond 
to results obtained with a $spd$ basis and the hcp and bcc lattices for 
Ti and W, respectively. In the case of Ti two further curves are given
which correspond to the fcc lattice with a $spd$ basis (broken line) 
or a $sd$ basis (chain line). The Fermi level was set at zero energy.}
\label{Bethe_Ti_W}
\end{figure}

The tight-binding parameters, which include only nearest-neighbors 
interactions, used in these calculations are given
in Table \ref{table}. The Table reports data not only for the metals
taken as electrodes in the present work (Al and Au) but also
for two additional metals (Ti and W) commonly used in experiments 
and/or calculations.
All were obtained through fittings to the electronic
bulk band structures calculated  by
including second or even third nearest-neighbors interactions\cite{Pa86}. 
In the case of hexagonal-close-packed Ti we took as nearest-neighbors  
six out of plane and six in-plane neighbors,
as actual interatomic distances differ in less than a 2\%\cite{AM76}.
The densities of states on  bulk atoms are shown in Fig. (\ref{Bethe_Al_Au})
and (\ref{Bethe_Ti_W}). Although 
some of the features of the actual DOS are not reproduced (as it 
commonly occurs in the Bethe lattice approximation) the overall
results are satisfactory. The major discrepancy is noted for Ti since
in this metal the Fermi level in the crystalline case lies in a valley 
of the DOS\cite{Pa86}. In the calculations reported 
in this work the $sp$ basis was used for Al. 
We have described the electronic structure of Au 
by means of the $spd$ basis (see Fig. \ref{Bethe_Al_Au}
and Table \ref{table}). In the case of Au and Ti, 
the reduced $sd$ basis gives an excessively  narrow conduction band. 
Finally, we note that in the case of Ti there are no major
differences between the DOS for the fcc and hcp lattices.

\clearpage
\widetext 
        
\begin{table}
\caption{Tight-binding parameters (in Rydbergs) used in the calculation of
the density of states and selfenergies of the Bethe lattices for 
the metals commonly taken as electrodes, namely, aluminium, gold, 
titanium and tungsten (the last two not considered in the present work). 
Orbital on-site energies are represented by $e_i$ while 
nearest-neighbor interactions are
denoted by $v_i$. The electronic configuration taken in each
case are: Al-a (3s$^2$3p$^1$), Al-b (3s$^2$3p$^1$3d$^0$),
Au-a (5d$^{10}$6s$^1$), Au-b (5d$^{10}$6s$^1$6p$^0$),
Ti-a and Ti-c (3d$^2$4s$^2$4p$^0$), Ti-b (3d$^2$4s$^2$) and W 
(5d$^{4}$6s$^2$6p$^0$). Actual lattices are: 
aluminium and gold, face-centered-cubic, titanium, (a) 
hexagonal-close-packed and (b) and (c) face-centered-cubic,
and wolframium, body-centered-cubic. The parameters were
obtained by fitting the bulk electronic band structures 
given in {\protect \onlinecite{Pa86}}.  The Fermi level $\epsilon_F$ 
corresponding to these parameters is also given.}

\begin{tabular}{|c|c|c|c|c|c|c|c|c|}
\hline
 Parameter & Al-a  & Al-b  & Au-a & Au-b   & Ti-a  & Ti-b &  Ti-c  & W  \\
\hline
$e_{s}$           & 0.35512 &  0.50658 & 0.41996 & 0.51034 & 0.88100& 0.73609
& 1.07296& 0.61616 \\
$e_{p}$           & 0.88653 &  1.05686 &    -    & 1.28039 & 1.15042&    -
& 1.38106& 1.46304 \\
$e_{d_{xy}}$      &    -    &  1.73644 & 0.22963 & 0.27529 & 0.69449& 0.65165
& 0.69143& 1.12072 \\
$e_{d_{xz}}$      &    -    &  1.73644 & 0.22963 & 0.27529 & 0.68617& 0.65165
& 0.69143& 1.12072 \\
$e_{d_{yz}}$      &    -    &  1.73644 & 0.22963 & 0.27529 & 0.68617& 0.65165
& 0.69143& 1.12072 \\
$e_{d_{x^2-y^2}}$ &    -    &  1.64457 & 0.22930 & 0.25542 & 0.69449& 0.63735
& 0.67293& 1.04350 \\
$e_{d_{3z^2-r^2}}$&    -    &  1.64457 & 0.22930 & 0.25542 & 0.69828& 0.63735
& 0.67293& 1.04350 \\
\hline
$v_{ss\sigma}$  & -0.04852  & -0.06225 &-0.06682 &-0.06931 &-0.06809&-0.06353
& -0.07960&-0.07315 \\
$v_{sp\sigma}$  & -0.08296  & -0.08914 &    -    & 0.08543 & 0.07676&     -
& 0.11204&-0.00419 \\
$v_{sd\sigma}$  &     -     &  0.08741 &-0.03868 &-0.05282 & 0.04883&-0.04223
& -0.04567&-0.07323 \\
$v_{pp\sigma}$  &  0.19317  &  0.16491 &    -    & 0.17166 & 0.09883&    -
& 0.17587& 0.37460 \\
$v_{pp\pi}$     &  0.06339  & -0.00999 &    -    &-0.01084 &-0.01646&    -
& -0.00386& 0.08730 \\
$v_{pd\sigma}$  &     -     & -0.17352 &    -    &-0.09305 & 0.06692&    -
& -0.05580&-0.23996 \\
$v_{pd\pi}$     &     -     &  0.04472 &    -    & 0.01008 &-0.02718&    -
& 0.03158&-0.06160 \\
$v_{dd\sigma}$  &     -     & -0.16416 &-0.04391 &-0.04872 &-0.05211&-0.04337
& -0.04794&-0.18762 \\
$v_{dd\pi}$     &     -     &  0.07776 & 0.03367 & 0.02494 & 0.02862& 0.03907
& 0.03542&-0.06543 \\
$v_{dd\delta}$  &     -     & -0.01279 &-0.00874 &-0.00462 &-0.00603&-0.00462
& -0.00985& 0.06917 \\
\hline
$\epsilon_F$ & 6.98 & 8.24 &  6.84  &  7.16 &  7.83  &  8.02   & 7.89  & 10.45
\\ 
\end{tabular}
\label{table}
\end{table}

\end{document}